\documentclass[12pt,manuscript,psfig,epsfig,pdffig,superscriptaddress,floatfix,nofootinbib]{revtex4}
\usepackage[latin9]{inputenc}
\usepackage{amsmath}
\usepackage{amssymb}
\usepackage{graphicx}
\usepackage{epstopdf}
\usepackage{pstricks}
\usepackage{color}

\makeatletter
\@ifundefined{textcolor}{}
{%
	\definecolor{BLACK}{gray}{0}
	\definecolor{WHITE}{gray}{1}
	\definecolor{RED}{rgb}{1,0,0}
	\definecolor{GREEN}{rgb}{0,1,0}
	\definecolor{BLUE}{rgb}{0,0,1}
	\definecolor{CYAN}{cmyk}{1,0,0,0}
	\definecolor{MAGENTA}{cmyk}{0,1,0,0}
	\definecolor{YELLOW}{cmyk}{0,0,1,0}
}

\usepackage{dcolumn}
\usepackage{bm}
\usepackage{amsfonts}
\usepackage{cancel}

\makeatother

\begin{document}
	
\title{Amplitude of $H \to \gamma Z$ process via one W loop in unitary gauge \\
 (I. Details of calculation with Dyson scheme)}


\author{Shi-Yuan Li}
\affiliation{Institute  of Theoretical Physics, Shandong University, Jinan 250100, P. R. China}







\begin{abstract}
 Decay amplitude of $ H \to \gamma Z $ process
via one W loop in the unitary gauge is presented.           %
      The divergent integrals including  those of high divergence orders typical of unitary gauge 
      are  arranged to cancel to get the  electromagnetic $U(1)$ gauge invariant  finite result, hence no contribution to the  renormalization  constant of $Z\gamma$-mixing in this 1-loop subprocess. 
   For the calculation of the Feynman diagrams employing the   Feynman rules, all the integrations of   the  propagator momenta and  all the $\delta$-functions representing the 4-momentum conservation of every vertex
 are retained    in the beginning. Therefore,  the  ambiguity of  setting independent  loop momentum for  divergences worse than logarithmic does not exist, and shift of integrated variable in such  divergent integrals   is eschewed.  The calculation are done in 4-dimension Minkowski momentum space without the aid of any regularization. The correct treatment on the  surface terms for the quadratic and logarithmic tensor integral  is one of  the key points.

This part I is devoted to the calculation details and the indications from the key surface terms.  Comparing with other gauge(s) and complete results for  $ H \to \gamma Z $ are left for part II.
 %
	
\end{abstract}
\date{\today}
\maketitle

\section{Introduction}\label{sec1}
The Glashow-Weinberg-Salam 
    electroweak (EW) theory is a SU(2)$\times$U(1) Yang-Mills gauge field theory, with the gauge symmetry 'broken' by  a scalar field via the Englert-Brout-Higgs-Guralnik-Hagen-Kibble Mechanism and the scalar field coupling  the fermion field in Yukawa style provides the mass term which distinguishes  various  SU(2)-doublet fermions. This theory  has been confirmed
from experiment in the sense that  the massive $W^\pm$ $Z$ particles v.s. the massless photon, and a 'remaining' neutral scalar particle which is generally referred to
 as the Higgs or the 'God' particle, all are well measured.
In general, a realistic calculation of the S-matrix or scattering/decay amplitude employing the quantized  field theory
of the standard model need to fix a specific gauge and it is adopted that the physical result  should be  independent from the choice (artificial rather than by nature) of the gauge.  However, recently, 
 careful revisit on the $H \to \gamma \gamma$ decay width in the unitary gauge and the $R_{\xi}$ gauge \cite{Wu:2017rxt,Wu:2016nqf,Gastmans:2015vyh,Gastmans:2011ks,Gastmans:2011wh} would like to imply some paradox.
 For a review  and remarks  on the uncertainties in this paradox, see e.g., \cite{Duch:2020was}.
  In all ways
   this paradox calls for  calculations in the unitary gauge for loop diagrams  to be extensively studied.   
   Many topics have been suggested  \cite{Wu:2017rxt}, and one of them is
 the  $H \to \gamma Z$ process via one W loop. Though less experimentally significant  \cite{Aad:2014fia, Chatrchyan:2013vaa}, 
  it can also be an important example to investigate
   in  the unitary gauge and  the $R_\xi$ gauge to gain insights for the paradox.
   %
 %

It is well known that unitary gauge can be taken  as  a limit of the general $R_\xi$
gauge  (but can be defined independently \cite{wein73}) that does not commute with the loop integrations \cite{Wu:2017rxt},  so that a lot of care
has to be taken when applied to loop calculations.  In the above mentioned gauge (non)invariance paradox, several uncertainties could arise from the  (maybe) non-commutability of  various limitations \cite{Duch:2020was}. Besides, high  divergence order integrals are one of the difficulties.
 Similar as  $H \to \gamma \gamma$ process via W loop,
  $H \to \gamma Z$ process via one W loop in the unitary gauge also has many  high   divergence order integrals  for each single diagrams  and  should properly cancel,  or else one can not get the correct  result, either not possible to make the comparison with results from other gauges. 
  Terms proportional   to $ M_Z^2$, not encountered in  $H \to \gamma \gamma$ process, cause   new difficulties.
  The purpose of this paper is to apply the experiences obtained from the investigation on the $H \to \gamma \gamma$ process via one W loop \cite{Li:2017hnv}, i.e.,
   without the setting of independent integrated loop momentum in  the beginning and  eschewing the shift of integrated variable for high order divergences,  to provide the  systematic framework to give the finite  and electromagnetic U(1) gauge invariant  amplitude of the $H \to \gamma Z$ process via one W loop for further study.
  %
   %

%
%
For any diagram whose divergence order higher than logarithmic, to shift the integral momentum can lead to extra terms with lower divergence (or finite).
In such case,   the proper set of diagrams with correct inter-relations of the  loop momenta must be treated together
 to get the correct result, as pointed by \cite{Gastmans:2011ks,Gastmans:2011wh} (in the following we refer to these two papers and works therein as GWW). Only a   part   of  diagrams of the set shifting the momenta will change the result. This  problem can  be solved by the the original Dyson formulation, 'Dyson scheme' as called in \cite{Li:2017hnv}, without the ambiguity of  setting independent  loop momentum in the beginning, and shift of integrated variable in high divergence order integrals can be, {\it and  is,}  eschewed. Correspondingly,  our cancellation of divergences are all at integral level rather than integrand level. 
 However,  a wise setting of independent integrated momentum as GWW  or employing  Dyson scheme \cite{Li:2017hnv} is     inadequate to definitely  determine the final result in the unitary gauge  
   because of the presence of the surface term for the reduction of  the divergent tensor integral.  In $H \to \gamma Z$ process,
   the   same logarithmic divergent tensor integral appears as in $H \to \gamma \gamma$ \cite{Li:2017hnv};  but there is also the new  quadratic ones, especially for terms proportional to $M_Z^2$,  which will be investigated  in this paper.
   %
 %


The considerations and key points of employing Dyson scheme are   as following:

To think  that a high  divergence order  ($>$ logartithmic)  integral  from the Feynman diagram is changed when shifting the integrated momentum,  one  inevitably raises the question what is the 'original' expression/value to be changed? There may not exist the 'original' one for a single diagram, once considering that different Feynman diagrams are related and hence   the loop momenta (\`{a} la GWW). However, one can have the definiteness starting  from the original form derived from the perturbative expansion of the S-matrix according  to the standard  Dyson-Wick procedure \cite{Dyson:1949ha}, which is integrations on space-time at each perturbative  order. Once taking these space-time integrations \footnote{leaving out the integrations on momenta from each propagators; this in fact  exchanging order of integration, between the phase space and configuration space.}, we get  $\delta$ functions, one  for each vertex, relating all the momenta of the propagators with energy-momentum conservation \cite{Dyson:1949ha}. If we start  from such a form for each diagram, without integrating the propagator momenta and  $\delta$ functions, there will be no indefiniteness, or ambiguity to set independent integral momentum.
This corresponds to that the momentum space Feynman rules   are slightly modified \footnote{in fact 'recovered', see the classical paper of Dyson \cite{Dyson:1949ha}, especially its Eq. (20) and discussions before and after it.} as:  Any propagator with momentum $q$ has an extra $[\int \frac{d^4 q}{(2\pi)^4}]$ 'operator',  i.e., should this integration on $q$ to be done in the calculation of the Feynman diagram;
  any vertex has an extra factor $ (2\pi)^4 \delta(\sum_i q_i)$, with  $q_i$,  each momentum of all the propagators attaching  the vertex,    incoming.
In this paper we adopt this way to write the amplitude corresponding to each Feynman diagram for calculation.  The method has been proved to be valid and feasible  based on our investigation on the   $H \to \gamma \gamma$ process via one W loop in the unitary gauge \cite{Li:2017hnv}. Calculation in this way can also help us to eschew the shift of integrated variables in high order divergences.
For the present $\gamma Z$ case in unitary gauge this seems the only practical way. In \cite{Wu:2017rxt}, to eliminate uncertainties, the authors suggested to calculate the difference between unitary gauge and $R_\xi$ gauge
by first to calculate the difference of the integrand and then to do the loop integration. Since the difference of the integrand  still
 leads to high order divergences, the choice of integrated momenta has also to be treated in the above suggested way to eliminated ambiguity.
There is another advantage in applying the Dyson scheme when treating the surface term for the divergent tensor integral reduction. Since the    integrated momentum is  just the one in  the originally-defined Feynman propagator, its natural physical boundary condition can be used to determine the surface term (to be zero).

In the following section 2, we illustrate the details of the calculations in the way mentioned above.
The result is finite, $U_{EM}(1)$ invariant, without the need of the  Dyson subtraction with the correct treatment on the logarithmic and quadratic surface terms. No regularization is introduced and all calculations are done in the 4-dimension Minkowski momentum space. During the calculation, only real convergence  or real logarithmic divergence need not to eschew shift.   As is also drawn attention by  \cite{Li:2017hnv}, especially for terms with odd power of momentum (fermion propagator is another example), it has long been noticed that the real divergence can be worse than simple power counting once all possible ways  for momentum  to infinity  considered---this is just the
condition when we do not beg for regularization.

The physical implication of the finite terms proportional to $M_Z^2/M^{4}$ after all divergences from quadratic to logarithmic cancelled is very interesting as well. In $\gamma \gamma$ case, the $M^{-2}$ term not zero is explained because of the goldstone effects, which is equivalently exposed in unitary gauge by the corresponding terms in W propagator (and going to the final result once the surface term of logarithmic tensor reduction and cancellation is properly treated).  In the $\gamma Z$ case, the Z particle includes transverse as well as longitudinal components, which can expose more longitudinal (goldstone) effects from the W propagator. This is true because of the extra effects proportional to $M^2_Z$.  


Besides the success of calculating these H decay channels, this way also shows  power to discuss other important issues. For example, in \cite{Bao:2021byx} we demonstrate that it is very easy to
 obtain both the vector  current and axial vector current conservations at the same time via this Dyson scheme  \footnote{It has been long recognized, both current conservations can be obtained at the same time  via setting 'the most symmetric loop momentum' as in \cite{Wu:2017rxt,Wu:2016nqf,Gastmans:2015vyh,Gastmans:2011ks,Gastmans:2011wh}. The author thanks Prof. T.T. Wu for informing this fact.}, contrary to the Bell-Jackiw claim. From such investigations, one also recognizes the important
 r\^{o}le of the infinite momentum surface integrals. So we make some discussions on the  results,  the Dyson scheme, the surface terms, and the physical  implication on probing the structures and properties of space-time in section 3.

This part I is devoted to the calculation details to get the electromagnetic  U(1) gauge invariant result and physical indication from this calculation procedure, especially the the divergence cancelation from the  surface term.  Comparing with other gauge(s) and complete results for  $ H \to \gamma Z $ are left for part II.

\section{Calculation}
  \subsection{}
There are totally 5 Feynman diagrams.
Three are similar as those of the $H \to \gamma \gamma$ process, via the direct coupling of Higgs and W-boson loop. We will demonstrate the calculations in details.
The other two are via the direct 3-point coupling of Higgs and ZZ, while one  Z transiting to gamma via the W bubble (3-point vertices) and W tadpole (4-point vertex). However,  these latter two diagrams sum to be zero \cite{Wu:2017rxt}, and the details of the calculation will not be presented in this paper. In this calculation the same  surface term of the quadratic tensor integral   is  encountered, which will be investigated in details for the former three diagrams in the following \footnote{As a matter of fact, these diagrams are self-energy like diagrams, and by Lorentz and $U(1)_{EM}$ gauge invariant arguments should be proportional to $k_1^2 g_{\mu \alpha}-k_{1\mu}k_{1 \alpha}$ ($\mu$ the on shell photon index, $\alpha$  some dummy index), so are zero. But just  the same as the QED photon self-energy diagram, direct calculation is non-trivial. The superficial quadratic  latter also need the correct quadratic surface term to get the 'proper' form, which was regarded only available via a gauge invariant regularization. With the same argument, and also can be shown explicitly by the help of the  same  quadratic surface term, the fermion bubble of $\gamma Z$ also zero since $k_1$ on shell, i.e.,  $k_1^2=0, k_1 \cdot \epsilon =0$. The application to the fermion case  in fact is even more useful since here we can freely work in 4 dimension Minkowski momentum space without the ambiguity of $\gamma^5$.}. This zero result together with  the finite result of the former three diagrams show  in this 1-loop subprocess no contribution to the $Z\gamma$ mixing renormalization constant.

   Figure 1 showes the three diagrams  to be calculated in the following, $k_1$ is the 4-momentum of  $\gamma$.
 $k_2$ is the 4-momentum of the Z particle, $k_2^2=M_Z^2$, with  the  corresponding  polarization vectors $\epsilon_{Z\lambda}^\nu$,
$\lambda=1,2,3$, since  Z is massive. We still have $ k_{2\nu} \epsilon_{Z\lambda}^\nu=0, ~ \forall \lambda$ (see Eq. \ref{a1}).  There is also
an extra $\cot\theta_W$ factor for the
WWZ vertex, where $\theta_W$ is the Weinberg angle. 
  \begin{figure}[htb]
   	\centering
   	\begin{tabular}{cccccc}
   		\scalebox{0.3}[0.3]{\includegraphics{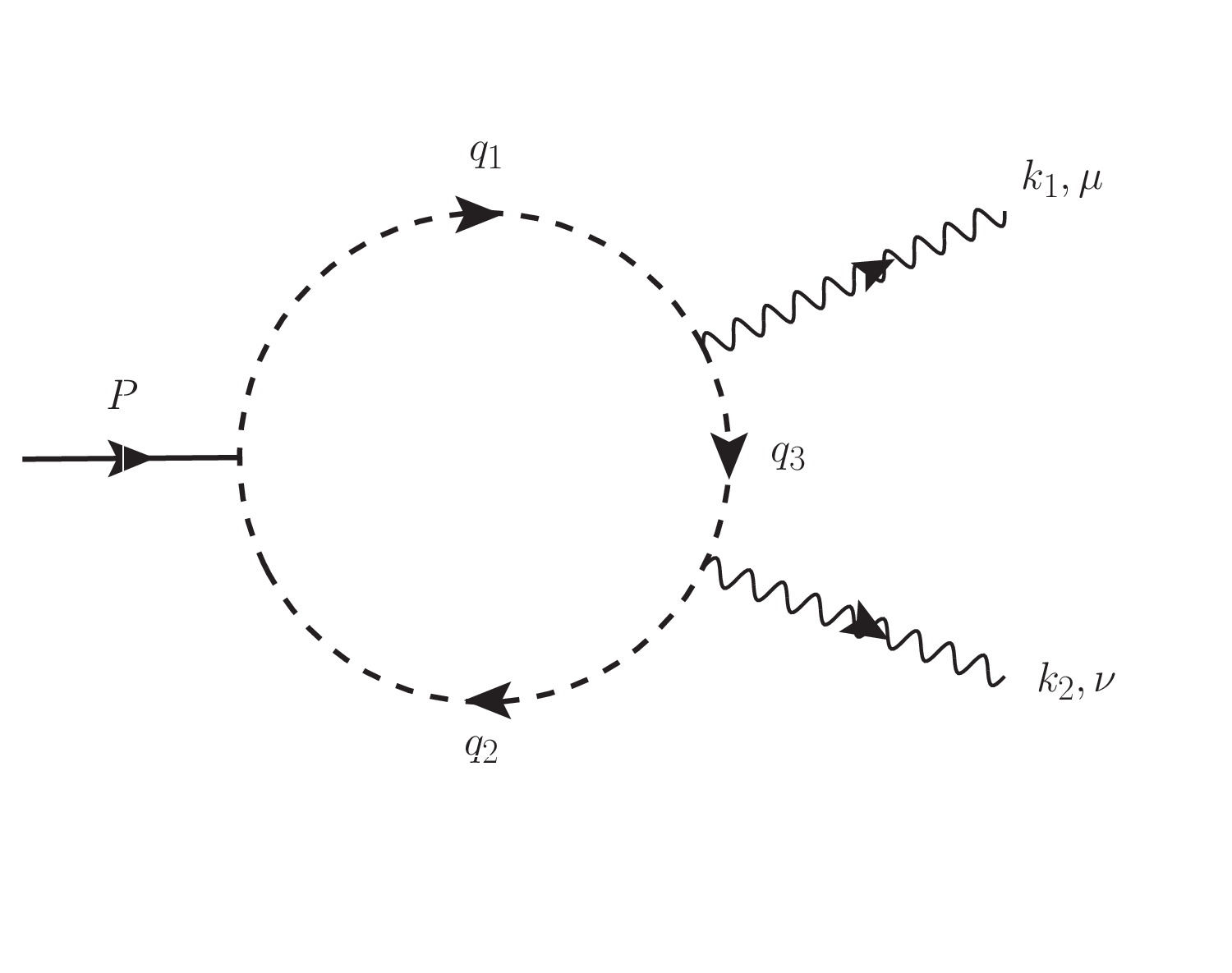}}&
   		\scalebox{0.3}[0.3]{\includegraphics{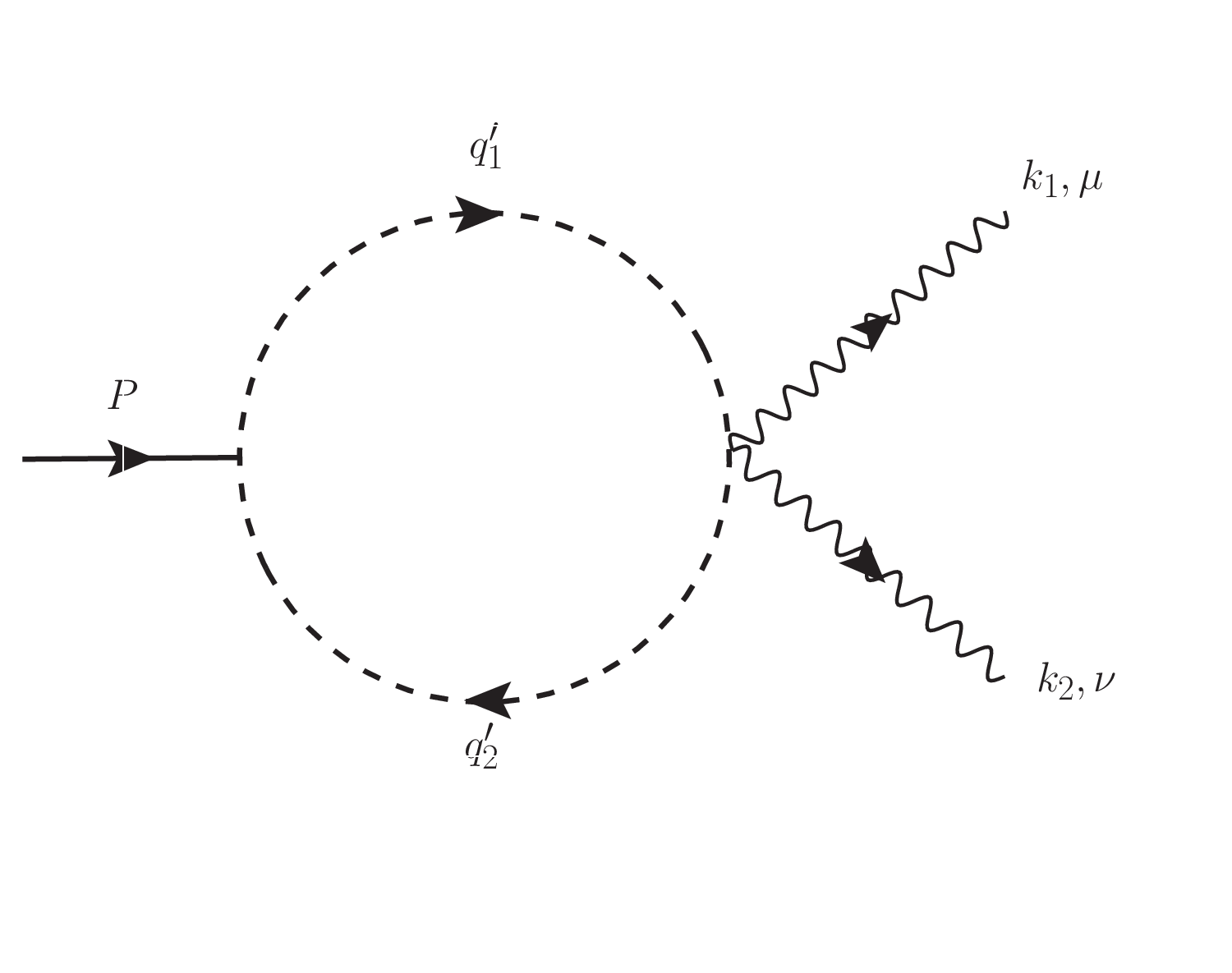}}&\\
   		{\scriptsize ($T_1$)}&{\scriptsize ($T_2$)}\\
   		\scalebox{0.33}[0.33]{\includegraphics{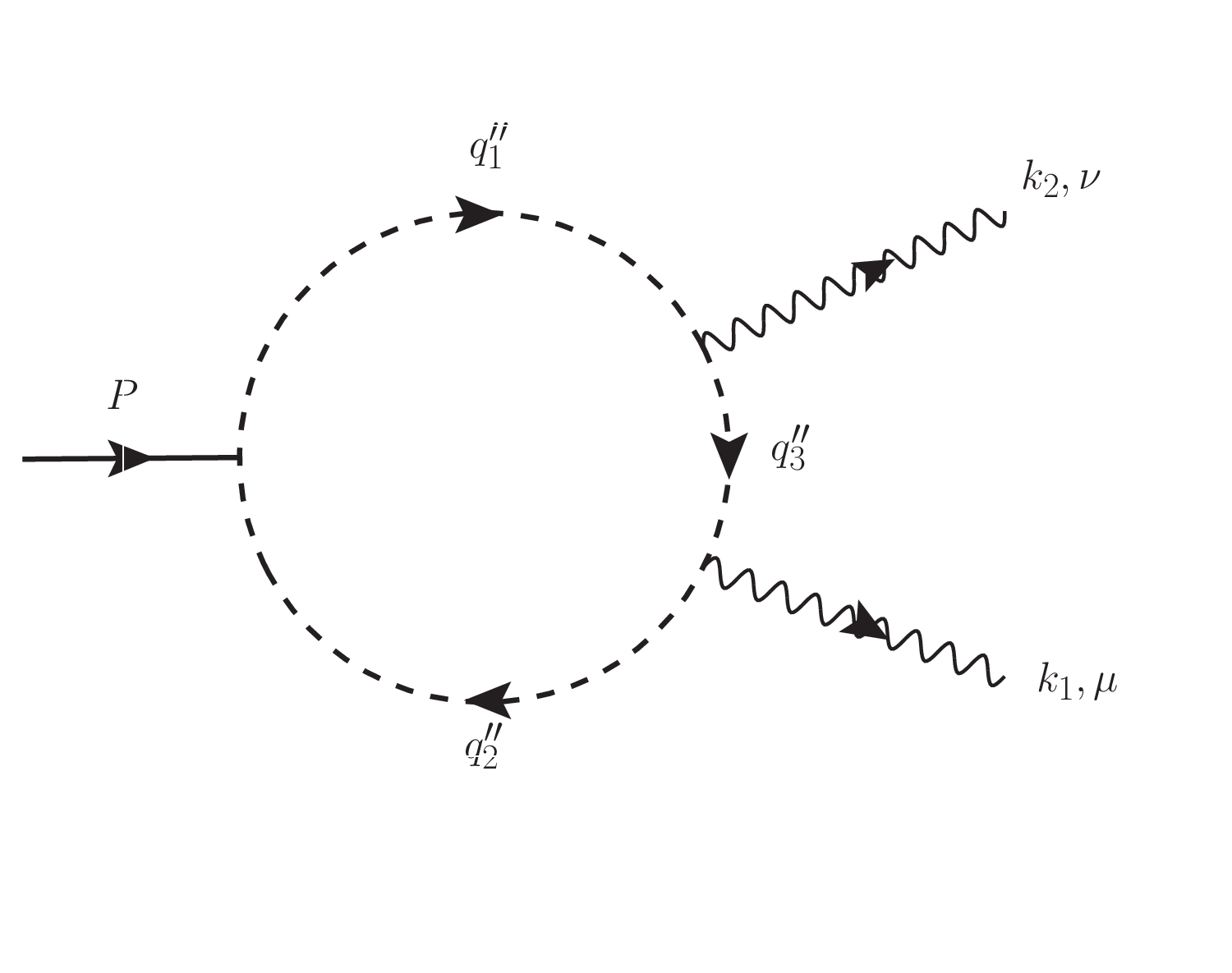}}&~\\
   		{\scriptsize ($T_3$)}& ~\\   		
   	\end{tabular}
   	\caption{The  one W loop Feynman diagrams with non-zero contribution to the process $H \to \gamma Z$. Taking the waving line with momentum $k_2$  represents the  Z particle with polarization vectors $\epsilon_{\lambda}^\nu$. In general, the inner integrated momenta should be considered as not correlated between different diagrams, so here we mark those of $T_2$ and $T_3$ with prime or double primes. In the manuscript, this is also implied though they are written as the same in the concrete step of derivation.  Since the  cancellation  is determined by the whole integral (with the integrand including the $\delta$ functions),  this way of writing dummy  variables does not lead to ambiguity and  is easily to be tracked. }\label{fd}
   \end{figure}




As convention, The S-matrix and T-matrix have the relation $S=I+i T$, and the matrix element between initial and final states
 $i T_{fi}=i (2\pi)^4 \delta(P_f-P_i) \mathfrak{M}_{fi}$
for the space-time displacement invariant case. 
Here  we keep all the  momenta respectively corresponding to  each propagator and hence all
$\delta$ functions respectively corresponding to each   vertex.  The one corresponding to  the initial-final state energy momentum conservation
is contained in these $\delta$ functions. After integrating over them, one will get the above form of T-matrix element with the $\mathfrak{M}_{fi}$
is the integration of the independent loop momenta only,  without the $\delta$ functions attached to the vertices.
This is the standard procedure in developing the Dyson-Wick perturbation theory in the interaction picture \cite{Dyson:1949ha}.  The four-momentum conservation
$\delta$ function attached to each of the vertices is the result of integration  of space-time variables in the perturbative expansion of the
S-matrix, and is the manifestation of space-time displacement invariance \footnote{This requires not only the boundary of space-time at infinity  trivial but also no  point or structure singular to 'block' the displacement. We conjecture this may be the condition to exchange the integration order of configuration space and momentum space, see footnote 1. We also would like to point out "no  point or structure singular to 'block' the displacement" guarantees the investigation on the surface term at infinity in momentum space \cite{Li:2017hnv}.}.
In the following, we do not integrate out the $\delta$ functions of each vertex until have to and is allowed to integrate out some of the momenta with corresponding $\delta$ functions
 (only for 'hidden' case, see the following).
So here we deal with the matrix elements $T_{fi} $ rather than $\mathfrak{M}_{fi}$:
\begin{eqnarray}
\label{t1}
 T_1 &= &  \frac{-ie^2gM\cot\theta_W}{(2 \pi)^4} \int d^4q_1 d^4q_2 d^4q_3 (2 \pi)^4 \delta(P-q_1+q_2) \delta(q_1-k_1-q_3)\delta(q_3-k_2-q_2)   \\
~& \times & (g_\alpha~^{\beta} - \frac{q_{1\alpha} q_1 ^{\beta}}{M^2} ) (g ^{\rho \sigma}-\frac{q_3 ^\rho q_3 ^\sigma}{M^2} ) (g^{\alpha \gamma} -\frac{q_2^\alpha q_2 ^\gamma}{M^2} )
 \frac{V_{\beta \mu \rho} (q_1,  -k_1, -q_3) ~ V_{\sigma \nu \gamma}(q_3, -k_2, -q_2)} {(q_1 ^2-M^2)(q_3 ^2-M^2)(q_2 ^2-M^2)}, \nonumber
\end{eqnarray}
\begin{eqnarray}
\label{t2}
 T_2 &= &  \frac{ie^2gM\cot\theta_W}{(2 \pi)^4} \int d^4q_1 d^4q_2  (2 \pi)^4 \delta(P-q_1+q_2) \delta(q_1-q_2-k_1-k_2)   \\
~& \times & (g_\alpha~^{\beta} - \frac{q_{1\alpha} q_1 ^{\beta}}{M^2} )  (g^{\alpha \gamma} -\frac{q_2^\alpha q_2 ^\gamma}{M^2} )
 \frac{2g_{\mu \nu} g_{\beta \gamma}-g_{\mu \beta} g_{\nu \gamma} -g_{\mu \gamma} g_{\nu \beta}} {(q_1 ^2-M^2)(q_2 ^2-M^2)}, \nonumber
\end{eqnarray}
\begin{eqnarray}
\label{t3}
 T_3 &= &  \frac{-ie^2gM\cot\theta_W}{(2 \pi)^4} \int d^4q_1 d^4q_2 d^4q_3 (2 \pi)^4 \delta(P-q_1+q_2) \delta(q_1-k_2-q_3)\delta(q_3-k_1-q_2)   \\
~& \times & (g_\alpha~^{\beta} - \frac{q_{1\alpha} q_1 ^{\beta}}{M^2} ) (g ^{\rho \sigma}-\frac{q_3 ^\rho q_3 ^\sigma}{M^2} ) (g^{\alpha \gamma} -\frac{q_2^\alpha q_2 ^\gamma}{M^2} )
 \frac{V_{\beta \nu \rho} (q_1,  -k_2, -q_3) ~ V_{\sigma \mu \gamma}(q_3, -k_1, -q_2)} {(q_1 ^2-M^2)(q_3 ^2-M^2)(q_2 ^2-M^2)}. \nonumber
\end{eqnarray}
 Here we do not explicitly write the matrix element subscript$~_{fi}$, and all the $\delta$ fuctions are understood as four-dimensional one,
i.e., $\delta(P_1-P_2) :=\delta^4 (P_1-P_2)$. As GWW, we also omit the polarization vector, so, e.g.,   $T_1$ should be understood as $T_{1\mu \nu}$.
In all this paper, we use M to represent the W mass.
   %
   %
  The relation between $T_1$ and $T_3$, i.e., $\mu \leftrightarrow \nu$, and $k_1 \leftrightarrow k_2$
is  clear to be read out. 


To better illustrate the calculation procedure, 
  we list the  formulae we use repeatedly  
  in Appendix A.  They correspond to Eqs.  (2.5-2.12) of GWW \cite{Gastmans:2011wh}  and can  go to GWW (2.5-2.12) by simply taking $M_Z=0$
  since there  all for photons.
   %
%
  Similarly as the $H \to \gamma \gamma$ process,  the property of 3-particle vertex, WW$\gamma$ or WWZ, which is named as Ward Identity (W.I.) in GWW, is the key role in the evaluation,
  because the extra terms $q^\alpha q^\beta/M^2$ in W propagator product with the vertex is typical of the unitary gauge.
  There is more subtle elements in considering these W.I.'s, since they demonstrate the special  relations of the various propagator momenta provided by the
  concrete dynamics in the standard mode. Furthermore, the equations (A9), (A10), and first and third terms of (A7),  (A8) are independent of the integrated momentum choice (the $(q_3^2-M^2)$ term reduces with one from the denominator, so in fact corresponding to a $g_{\mu\nu}$ term).

In the following, we investigate the terms according to their minus power of  M.  The $\frac{-ieg^2M\cos\theta_W}{(2 \pi)^4}$ factor will not explicitly written, and  all terms should multiply with this factor to get the
proper terms in the corresponding T amplitude of Eqs. (\ref{t1}-\ref{t3}). So when we mention a term, we do not take into account the overall M factor coming from the
HWW coupling except explicitly addressed.
Since there is still a WW$\gamma$ vertex,  i.e.,  $V_{\beta \mu \rho} (q_1,  -k_1, -q_3)$ of $T_1$, $V_{\sigma \mu \gamma}(q_3, -k_1, -q_2)$ of $T_3$, respectively, the $M^{-6}$ terms in $T_{1}$ and  $T_{3}$ are again zero as the $H \to \gamma \gamma$ process, according to Eq. (\ref{a9}).

 \subsection{}
For the $M^{-4}$ terms, in $T_{1}$ and  $T_{3}$, respectively, there are 3 ways of the combination of two of  three  $q_i^{\alpha_i} q_i^{\beta_i}/M^2$ ($i=1,2,3$) from the W propagators.
One combination is zero because of the W.I. of the WW$\gamma$ vertex, Eq. (\ref{a9}).  Another gives a $M_Z^2/M^4$ terms because of the W.I. of the WWZ vertex,  Eq. (\ref{a10}).
The third corresponds to those of the  $H \to \gamma \gamma$ process, gives likely terms besides extra $M_Z^2/M^4$ terms.

  Those from $T_1$:
\begin{eqnarray}
T''_{11}&=&\frac{1}{M^4} \int d^4q_1 d^4q_2 d^4q_3 (2 \pi)^4 \delta(P-q_1+q_2) \delta(q_1-k_1-q_3)\delta(q_3-k_2-q_2)\\
~ & \times &  q_{1\alpha} q_1 ^{\beta} q_3^\rho q_3^\sigma g ^{\alpha \gamma} \frac{V_{\beta \mu \rho} (q_1,  -k_1, -q_3) ~ V_{\sigma \nu \gamma}(q_3, -k_2, -q_2)} {(q_1 ^2-M^2)(q_3 ^2-M^2)(q_2 ^2-M^2)}=0, \nonumber
\end{eqnarray}
\begin{eqnarray}
T'_{11}&=&\frac{1}{M^4} \int d^4q_1 d^4q_2 d^4q_3 (2 \pi)^4 \delta(P-q_1+q_2) \delta(q_1-k_1-q_3)\delta(q_3-k_2-q_2)\\
~ & \times &  g_{\alpha}~ ^{\beta} q_3^\rho q_3^\sigma q_2 ^{\alpha} q_2 ^{\gamma} \frac{V_{\beta \mu \rho} (q_1,  -k_1, -q_3) ~ V_{\sigma \nu \gamma}(q_3, -k_2, -q_2)} {(q_1 ^2-M^2)(q_3 ^2-M^2)(q_2 ^2-M^2)} \nonumber \\
   &=&\frac{M_Z^2}{M^4} \int d^4q_1 d^4q_2 d^4q_3 (2 \pi)^4 \delta(P-q_1+q_2) \delta(q_1-k_1-q_3)\delta(q_3-k_2-q_2) \nonumber \\
~ & \times &   \frac{(-q_1^2 q_{2\mu} q_{3\nu}+q_1\cdot q_2 q_{1\mu} q_{3\nu}) } {(q_1 ^2-M^2)(q_3 ^2-M^2)(q_2 ^2-M^2)} \nonumber,
\end{eqnarray}
employing the corresponding W.I.'s of Appendix A.  Obviously, $T'_{11}=0$ if $M_Z=0$, which is consistent with the $H \to \gamma \gamma$ case.
\begin{eqnarray}
\label{t11}
T_{11}&=&\frac{1}{M^4} \int d^4q_1 d^4q_2 d^4q_3 (2 \pi)^4 \delta(P-q_1+q_2) \delta(q_1-k_1-q_3)\delta(q_3-k_2-q_2)\\
~ & \times &  q_{1\alpha} q_1 ^{\beta} g^{\rho \sigma} q_2 ^{\alpha} q_2 ^{\gamma} \frac{V_{\beta \mu \rho} (q_1,  -k_1, -q_3) ~ V_{\sigma \nu \gamma}(q_3, -k_2, -q_2)} {(q_1 ^2-M^2)(q_3 ^2-M^2)(q_2 ^2-M^2)}. \nonumber
\end{eqnarray}
Now since $q_1 ^{\beta} V_{\beta \mu \rho} (q_1,  -k_1, -q_3)= (q_3^2-M^2) g_{\mu \rho}-q_{3\mu}q_{3\rho}+M^2 g_{\mu \rho}$, according to (\ref{a7}),
 $T_{11}=T_{111}+T_{112}+T_{113}$.  For this kind of  step we  employ the W.I. for WW$\gamma$ (A7) in priority than  WWZ vertex (A8) because it straightforwardly
 gives the proper minus  power of M for each term.
 \begin{eqnarray}
\label{t113}
T_{113}&=&\frac{M^2}{M^4} \int d^4q_1 d^4q_2 d^4q_3 (2 \pi)^4 \delta(P-q_1+q_2) \delta(q_1-k_1-q_3)\delta(q_3-k_2-q_2)\\
~ & \times &  q_{1} \cdot  q_2  \frac{V_{\mu \nu \gamma}(q_3, -k_2, -q_2)q_2 ^{\gamma}} {(q_1 ^2-M^2)(q_3 ^2-M^2)(q_2 ^2-M^2)} \nonumber
\end{eqnarray}
is obviously $M^{-2}$ term and to be discussed with other $M^{-2}$ terms later.
\begin{eqnarray}
\label{t112}
T_{112}&=&\frac{1}{M^4} \int d^4q_1 d^4q_2 d^4q_3 (2 \pi)^4 \delta(P-q_1+q_2) \delta(q_1-k_1-q_3)\delta(q_3-k_2-q_2)\\
~ & \times & q_{1} \cdot  q_2 q_{3\mu}  \frac{q_3^\sigma V_{\sigma \nu \gamma}(q_3, -k_2, -q_2)(-q_2 ^{\gamma})} {(q_1 ^2-M^2)(q_3 ^2-M^2)(q_2 ^2-M^2)} \nonumber \\
  &=&\frac{M_Z^2}{M^4} \int d^4q_1 d^4q_2 d^4q_3 (2 \pi)^4 \delta(P-q_1+q_2) \delta(q_1-k_1-q_3)\delta(q_3-k_2-q_2) \nonumber \\
~ & \times &   \frac{  q_{1} \cdot  q_2 q_{3\mu} q_{3\nu} } {(q_1 ^2-M^2)(q_3 ^2-M^2)(q_2 ^2-M^2)}, \nonumber
\end{eqnarray}
 \begin{eqnarray}
 \label{t111}
T_{111} &=&\frac{1}{M^4} \int d^4q_1 d^4q_2 d^4q_3 (2 \pi)^4 \delta(P-q_1+q_2) \delta(q_1-k_1-q_3)\delta(q_3-k_2-q_2)\\
~ & \times &  q_1 \cdot  q_2    \frac{V_{\mu \nu \gamma}(q_3, -k_2, -q_2)q_2 ^{\gamma}} {(q_1 ^2-M^2)(q_2 ^2-M^2)} \nonumber \\
  &=&\frac{1}{M^4} \int d^4q_1 d^4q_2 d^4q_3 (2 \pi)^4 \delta(P-q_1+q_2) \delta(q_1-k_1-q_3)\delta(q_3-k_2-q_2) \nonumber \\
~ & \times &  q_1 \cdot  q_2  \frac{ q_1 \cdot q_2 g_{\mu \nu} -q_{1\mu} q_{2\nu} +(q_1\cdot k_2-q_2\cdot k_1-k_1\cdot k_2) g_{\mu\nu} -M_Z^2 g_{\mu \nu} } {(q_1 ^2-M^2)(q_2 ^2-M^2)}. \nonumber 
\end{eqnarray}
Again to employ the corresponding W.I. of these two kinds of 3-particle vertices.  
In the last  step of Eq. (\ref{t111}), we also take into account that, $\int dx f(x) \delta(x-a)=\int dx f(a) \delta(x-a)$, to use the relation $q_3=q_1-k_1=q_2+k_2$,
to get $$ (q_3^2 g_{\mu \nu}-q_{3\mu}q_{3\nu} )= q_1 \cdot q_2 g_{\mu \nu} -q_{1\mu} q_{2\nu} +(q_1\cdot k_2-q_2\cdot k_1-k_1\cdot k_2) g_{\mu\nu}.$$
In fact, all the Ward identities for the 3-boson vertex we use here also have employed the energy momentum conservation at the vertex.

We write
\begin{eqnarray}
T_{11B}&=& T'_{11}+T_{112}\\
 &=&\frac{ M_Z^2}{M^4} \int d^4q_1 d^4q_2 d^4q_3 (2 \pi)^4 \delta(P-q_1+q_2) \delta(q_1-k_1-q_3)\delta(q_3-k_2-q_2) \nonumber \\
~ & \times &   \frac{(-q_1^2 q_{2\mu} q_{2\nu}+ 2 q_1\cdot q_2 q_{3\mu} q_{3\nu})} {(q_1 ^2-M^2)(q_3 ^2-M^2)(q_2 ^2-M^2)} \nonumber,
\end{eqnarray}
so all the $M^{-4}$ term from $T_1$ are now $T_{11B}$ and $T_{111}$, with the  4th and 3th order divergences only in $T_{111}$ (formally similar as those in $H\to \gamma \gamma$).

Now we come to $T_3$,  from external variables, its relation with $T_1$ is $\mu \leftrightarrow \nu$, and $k_1 \leftrightarrow k_2$; from
 internal momenta, it is $q_1 \leftrightarrow -q_2$ and $q_3 \leftrightarrow -q_3$.
For easy to investigate, we now deal with it separately, according to the same thread of $T_1$.  Similarly for the three ways of combination,
$T''_{31}=0$, and
\begin{eqnarray}
T'_{31}&=&\frac{M_Z^2}{M^4} \int d^4q_1 d^4q_2 d^4q_3 (2 \pi)^4 \delta(P-q_1+q_2) \delta(q_1-k_2-q_3)\delta(q_3-k_1-q_2)  \\
~ & \times &   \frac{(-q_2^2 q_{1\mu} q_{1\nu}+q_1\cdot q_2 q_{3\mu} q_{3\nu}) } {(q_1 ^2-M^2)(q_3 ^2-M^2)(q_2 ^2-M^2)} \nonumber.
\end{eqnarray}
 For $T_{31}$, we also have $T_{31}=T_{311}+T_{312}+T_{313}$, with
\begin{eqnarray}
\label{t313}
T_{313}&=&\frac{M^2}{M^4} \int d^4q_1 d^4q_2 d^4q_3 (2 \pi)^4 \delta(P-q_1+q_2) \delta(q_1-k_2-q_3)\delta(q_3-k_1-q_2)\\
~ & \times &   q_{1} \cdot  q_2  \frac{ q_1 ^\beta V_{\beta \nu \mu  }(q_1, -k_2, -q_3)} {(q_1 ^2-M^2)(q_3 ^2-M^2)(q_2 ^2-M^2)} \nonumber
\end{eqnarray}
is obviously $M^{-2}$ term and to be discussed with other $M^{-2}$ terms later.
\begin{eqnarray}
T_{312}&=&\frac{ M_Z^2}{M^4} \int d^4q_1 d^4q_2 d^4q_3 (2 \pi)^4 \delta(P-q_1+q_2) \delta(q_1-k_2-q_3)\delta(q_3-k_1-q_2)  \\
~ & \times &   \frac{  q_{1} \cdot  q_2 q_{3\mu} q_{3\nu} } {(q_1 ^2-M^2)(q_3 ^2-M^2)(q_2 ^2-M^2)}, \nonumber
\end{eqnarray}
  \begin{eqnarray}
T_{311}&=&\frac{1}{M^4} \int d^4q_1 d^4q_2 d^4q_3 (2 \pi)^4 \delta(P-q_1+q_2) \delta(q_1-k_2-q_3)\delta(q_3-k_1-q_2)  \\
~ & \times &  q_1 \cdot  q_2  \frac{ q_1 \cdot q_2 g_{\mu \nu} -q_{1\nu} q_{2\mu} +(q_1\cdot k_1-q_2\cdot k_2-k_1\cdot k_2) g_{\mu\nu} -M_Z^2 g_{\mu \nu} } {(q_1 ^2-M^2)(q_2 ^2-M^2)}. \nonumber 
\end{eqnarray}
We write
\begin{eqnarray}
T_{31B}&=& T'_{31}+T_{312}\\
 &=&\frac{M_Z^2}{M^4} \int d^4q_1 d^4q_2 d^4q_3 (2 \pi)^4 \delta(P-q_1+q_2) \delta(q_1-k_2-q_3)\delta(q_3-k_1-q_2) \nonumber \\
~ & \times &   \frac{(-q_2^2 q_{1\mu} q_{1\nu}+ 2 q_1\cdot q_2 q_{3\mu} q_{3\nu}) } {(q_1 ^2-M^2)(q_3 ^2-M^2)(q_2 ^2-M^2)} \nonumber,
\end{eqnarray}
 again all the $M^{-4}$ term from $T_3$ are now $T_{31B}$ and $T_{311}$, with the  4th and 3th order divergences only in   $T_{311}$. 
    $T_{111}$ and $T_{311}$  can be combined, since for both of them $q_3$ only appear in the $\delta $ functions and can be integrated out.
  Then     the  4th and 3th order divergences cancel after summing with those of $T_2$
   and only quadratic one proportional to $M_Z^2/M^4$ left (zero for $M_Z=0$).
 \begin{eqnarray}
T_{111} +T_{311}&=&\frac{1}{M^4} \int d^4q_1 d^4q_2  (2 \pi)^4 \delta(P-q_1+q_2) \delta(q_1-q_2-k_1-k_2)\\
~ & \times &    q_1 \cdot  q_2 ~~ \frac{2 q_1 \cdot q_2 g_{\mu \nu} -q_{1\mu} q_{2\nu}-q_{2\mu} q_{1\nu} -M_Z^2 g_{\mu \nu}} {(q_1 ^2-M^2)(q_2 ^2-M^2)},  \nonumber 
\end{eqnarray}
since $ (q_1\cdot k_2-q_2\cdot k_1-k_1\cdot k_2) g_{\mu\nu}+ (q_1\cdot k_1-q_2\cdot k_2-k_1\cdot k_2) g_{\mu\nu}=k_2^2 g_{\mu\nu}= M_Z^2 g_{\mu\nu}$, takeing now $q_1-q_2=k_1+k_2$ from the $\delta$ function.
 The $M^{-4}$ term in $T_2$ is 
 \begin{eqnarray}
T_{21}&=&\frac{-1}{M^4} \int d^4q_1 d^4q_2  (2 \pi)^4 \delta(P-q_1+q_2) \delta(q_1-q_2-k_1-k_2)\\
~ & \times &    q_1 \cdot  q_2 ~~ \frac{2 q_1 \cdot q_2 g_{\mu \nu} -q_{1\mu} q_{2\nu}-q_{2\mu} q_{1\nu} } {(q_1 ^2-M^2)(q_2 ^2-M^2)},  \nonumber 
\end{eqnarray}
so $T_{111} +T_{311}+T_{21}=T_A$: 
  %
 \begin{eqnarray}
 \label{ta}
T_{A}&=&\frac{1}{M^4} \int d^4q_1 d^4q_2   (2 \pi)^4 \delta(P-q_1+q_2) \delta(q_1-q_2-k_1-k_2)
    \frac{-M_Z^2 } {(q_1 ^2-M^2)(q_2 ^2-M^2)} \\
    ~&\times & q_1 \cdot  q_2  g_{\mu \nu}, \nonumber \\
   ~ &=&\frac{M_Z^2}{M^4} \int d^4q_1 d^4q_2   (2 \pi)^4 \delta(P-q_1+q_2) \delta(q_1-q_2-k_1-k_2)
    \frac{1 } {(q_1 ^2-M^2)(q_2 ^2-M^2)}\nonumber \\
    ~&\times & (\frac{(k_1+k_2)^2}{2}+(-\frac{(q_1^2-M^2)+(q_2^2-M^2)}{2}-M^2)) g_{\mu \nu}.\nonumber
\end{eqnarray}
All the uncancelled terms of $M^{-4}$ order, $T_A$, $T_{11B}$, $T_{31B}$,
are proportional to $M_Z^2$, so  are zero when Z mass goes to zero.
 Since $M_Z$ is a constant parameter in the dynamics,   
 the result is not because of the in-proper choice of the integrated
variables; as a matter of fact,     all the above derivation is independent from any special choice of the integrated variable.
In the integral $T_A$, the $q_3$ does not appear (or only appearing in the $\delta$ functions and is integrated out). This  case, which has appeared above,  is called 'hidden' in the following.

 We have
\begin{eqnarray}
\label{t11b}
T_{11B} &=&\frac{M_Z^2}{M^4} \int d^4q_1 d^4q_2 d^4q_3  (2 \pi)^4 \delta(P-q_1+q_2) \delta(q_1-k_1-q_3) \delta(q_3-k_2-q_2) \\
    ~&\times &\frac{1 } {(q_1 ^2-M^2)(q_2 ^2-M^2) (q_3^2-M^2)} \nonumber \\
    ~&\times & (-(k_1+k_2)^2 q_{3\mu}q_{3\nu}+q_2^2 q_{3\mu}q_{3\nu}+(q_3^2-M^2)k_{2\mu}q_{2\nu}+M^2 k_{2\mu}q_{2\nu}+2 k_1\cdot q_3k_{2\mu}q_{3\nu});\nonumber
\end{eqnarray}
\begin{eqnarray}
\label{t31b}
T_{31B}&=&\frac{M_Z^2}{M^4} \int d^4q_1 d^4q_2 d^4q_3  (2 \pi)^4 \delta(P-q_1+q_2) \delta(q_1-k_2-q_3) \delta(q_3-k_1-q_2) \\
    ~&\times &\frac{1 } {(q_1 ^2-M^2)(q_2 ^2-M^2) (q_3^2-M^2)} \nonumber \\
     ~&\times & (-(k_1+k_2)^2 q_{3\mu}q_{3\nu}+q_1^2 q_{3\mu}q_{3\nu}-(q_3^2-M^2)k_{2\mu}q_{1\nu}-M^2 k_{2\mu}q_{1\nu}+2 k_1\cdot q_3k_{2\mu}q_{3\nu}).\nonumber
\end{eqnarray}
In the above equations, $(q_1-q_2)=(k_1+k_2)$ is used. In the  last line of each of Eqs. (\ref{ta},\ref{t11b}, \ref{t31b}),  we have decomposed them into various terms with various orders of divergences.

Before discussing the $M_Z^2/M^4$ terms, we first collect the uncancelled finite $M_Z^2/M^2$ terms appearing in above equations for further investigation.
The ones directly read from the Eqs. (\ref{t11b}) and (\ref{t31b}) respectively  are the fourth term in last line of each 
\begin{eqnarray}
\label{t11b1}
T_{11B1}&=&\frac{M_Z^2}{M^2} \int d^4q_1 d^4q_2 d^4q_3  (2 \pi)^4 \delta(P-q_1+q_2) \delta(q_1-k_1-q_3) \delta(q_3-k_2-q_2)\\
    ~&\times &\frac{ k_{2\mu}q_{2\nu}} {(q_1 ^2-M^2)(q_2 ^2-M^2) (q_3^2-M^2)} , \nonumber
\end{eqnarray}
\begin{eqnarray}
\label{t31b1}
T_{31B1}&=&\frac{M_Z^2}{M^2} \int d^4q_1 d^4q_2 d^4q_3  (2 \pi)^4 \delta(P-q_1+q_2) \delta(q_1-k_2-q_3) \delta(q_3-k_1-q_2)\\
    ~&\times &\frac{- k_{2\mu}q_{1\nu} } {(q_1 ^2-M^2)(q_2 ^2-M^2) (q_3^2-M^2)} . \nonumber
\end{eqnarray}
The denominators are not the same, since different kinematic configuration.  Similar attention should be paid in the following.
These two terms (\ref{t11b1}, \ref{t31b1}) are latter in need to be combined to get the $U_{EM}(1)$ gauge invariant term. This very subtle fact is a signal of the self-consistency of the standard model.

The linear divergent terms in Eqs. (\ref{t11b}) and (\ref{t31b}) respectively  are the third  term in last line of each. 
 The ($q_3^2-M^2$) reduces
the corresponding factor in denominator, and then they are independent on $q_3$. Integrating on $q_3$ in both and they can combine and give
\begin{equation}
\label{bb1}
\frac{M_Z^2}{M^4} \int d^4q_1 d^4q_2   (2 \pi)^4 \delta(P-q_1+q_2) \delta(q_1-q_2-k_1-k_2)
    \frac{-1 } {(q_1 ^2-M^2)(q_2 ^2-M^2)} k_{2\mu} k_{1\nu}.
    \end{equation}
    Again $(q_1-q_2)=(k_1+k_2)$ is used.

Further, dividing Eq. (\ref{bb1}) by 2, each respectively recover a $(q_3^2-M^2)$ factor in the numerator and denominator,  recover a third $\delta $
function and integration on $q_3$ corresponding to $T_{1}$ and $T_{3}$. The $q_3^2$ term of numerator   cancels the logarithmic divergence of  fifth (last) term in last line
respectively of     Eqs. (\ref{t11b}) and (\ref{t31b}) (the remaining finite terms are in the following).
The remaining $M_Z^2/M^2$ terms are
\begin{eqnarray}
\label{t11b2}
T_{11B2}&=&\frac{M_Z^2}{ M^2} \int d^4q_1 d^4q_2 d^4q_3  (2 \pi)^4 \delta(P-q_1+q_2) \delta(q_1-k_1-q_3) \delta(q_3-k_2-q_2)\\
    ~&\times &\frac{k_{2\mu}k_{1\nu}/2 } {(q_1 ^2-M^2)(q_2 ^2-M^2) (q_3^2-M^2)} , \nonumber
\end{eqnarray}
\begin{eqnarray}
\label{t31b2}
T_{31B2}&=&\frac{M_Z^2}{ M^2} \int d^4q_1 d^4q_2 d^4q_3  (2 \pi)^4 \delta(P-q_1+q_2) \delta(q_1-k_2-q_3) \delta(q_3-k_1-q_2)\\
    ~&\times &\frac{k_{2\mu}k_{1\nu}/2 } {(q_1 ^2-M^2)(q_2 ^2-M^2) (q_3^2-M^2)} . \nonumber
\end{eqnarray}

Similar cancellation  as above is also done for first term of   last line
respectively of     Eqs. (\ref{t11b}) and (\ref{t31b}) with the  first term of last line of the equation of $T_A$ (the remaining finite terms are in the following),
and the remaining $M_Z^2/M^2$ terms from $T_A$ (coming from the $q_3^2-M^2$ factor for both numerator and denominator, recovering the $\int dq_3$ integration) are
\begin{eqnarray}
\label{t11b3}
T_{A11}&=&\frac{M_Z^2}{ M^2} \int d^4q_1 d^4q_2 d^4q_3  (2 \pi)^4 \delta(P-q_1+q_2) \delta(q_1-k_1-q_3) \delta(q_3-k_2-q_2)\\
    ~&\times &\frac{ -(k_1+k_2)^2 g_{\mu\nu}/4}{(q_1 ^2-M^2)(q_2 ^2-M^2) (q_3^2-M^2)} , \nonumber
\end{eqnarray}
\begin{eqnarray}
\label{t31b3}
T_{A31}&=&\frac{M_Z^2}{ M^2} \int d^4q_1 d^4q_2 d^4q_3  (2 \pi)^4 \delta(P-q_1+q_2) \delta(q_1-k_2-q_3) \delta(q_3-k_1-q_2)\\
    ~&\times &\frac{- (k_1+k_2)^2  g_{\mu\nu}/4 } {(q_1 ^2-M^2)(q_2 ^2-M^2) (q_3^2-M^2)} . \nonumber
\end{eqnarray}
     %
   %
     %
   In the above two logarithmic divergence cancellations,
   the nonzero finite $M_Z^2/M^4$ terms are
   \begin{eqnarray}
T_{11BF}&=&\frac{M_Z^2}{ M^4} \int d^4q_1 d^4q_2 d^4q_3  (2 \pi)^4 \delta(P-q_1+q_2) \delta(q_1-k_1-q_3) \delta(q_3-k_2-q_2)\\
    ~&\times &\frac{\frac{1}{4} (k_1+k_2)^2 q_3^2 g_{\mu\nu}-(k_1+k_2)^2 q_{3\mu}q_{3\nu}- \frac{1}{2}  q_3^2 k_{2\mu} k_{1\nu} +2 k_1\cdot q_3 k_{2 \mu} q_{3\nu}} {(q_1 ^2-M^2)(q_2 ^2-M^2) (q_3^2-M^2)} , \nonumber
\end{eqnarray}
\begin{eqnarray}
T_{31BF}&=&\frac{M_Z^2}{ M^4} \int d^4q_1 d^4q_2 d^4q_3  (2 \pi)^4 \delta(P-q_1+q_2) \delta(q_1-k_2-q_3) \delta(q_3-k_1-q_2)\\
    ~&\times &\frac{\frac{1}{4} (k_1+k_2)^2 q_3^2 g_{\mu\nu}-(k_1+k_2)^2 q_{3\mu}q_{3\nu}- \frac{1}{2}  q_3^2 k_{2\mu} k_{1\nu} +2 k_1\cdot q_3 k_{2 \mu} q_{3\nu} } {(q_1 ^2-M^2)(q_2 ^2-M^2) (q_3^2-M^2)} . \nonumber
\end{eqnarray}
For the above six remaining non zero finite terms, contrary to (\ref{t11b1}, \ref{t31b1}),   it can be shown that Eqs. (\ref{t11b2}) to (\ref{t31b3}) are cancelled  by those terms in $T_{11BF}$ and $T_{31BF}$ proportional to an extra $M^2$ factor.   Those remaining in $T_{11BF}$ and $T_{31BF}$  all  proportional to $(k_1\cdot k_2 g^{\mu \nu}-k_2^\mu k_1^\nu)M_Z^2/M^4$. This in fact only can be approved after the integration on Feynman-Schwinger parameters $x_1, x_2$, by which extra terms other than the gauge invariant one integrated to be zero.
The coefficient of   $(k_1\cdot k_2 g^{\mu \nu}-k_2^\mu k_1^\nu)M_Z^2/M^4$ is then
\begin{equation}
\frac{i}{(4 \pi)^2}\int dx_1 dx_2 dx_3 \delta(x_1+x_2+x_3-1)\frac{x_1x_2 (2 k_1\cdot k_2+M_Z^2)}{x_1x_2 2 k_1\cdot k_2-x_2^2M_Z^2+x_2M_Z^2-M^2}.
\end{equation}
The result is of the form
\begin{equation}
\frac{1}{2}+O(\frac{M_Z^2}{2 k_1\cdot k_2}) \cdot \cdot \cdot),
\end{equation}
which guarantee the non-zero $M_Z^2/M^4$ terms in the unitary gauge.
 %


This term  is the most important part of   the $U_{EM}(1)$ gauge invariant final result, in the sense crucial to comparing with the  result from other gauges (e.g., $R_\xi$ gauge). For feasibility to collect for the whole result, we mark it as R1.


Now the   terms including quadratically  divergence  to be considered are:
 \begin{eqnarray}
 \label{tcq}
      ~& &\frac{1}{M^4} \int d^4q_1 d^4q_2   (2 \pi)^4 \delta(P-q_1+q_2) \delta(q_1-q_2-k_1-k_2)
    \frac{M_Z^2 } {(q_1 ^2-M^2)(q_2 ^2-M^2)} \\
    ~&\times & (-\frac{(q_1^2-M^2)+(q_2^2-M^2)}{2}-M^2) g_{\mu \nu}, \nonumber
\end{eqnarray}
\begin{eqnarray}
\label{t11bq}
 ~& &\frac{1}{M^4} \int d^4q_1 d^4q_2 d^4q_3  (2 \pi)^4 \delta(P-q_1+q_2) \delta(q_1-k_1-q_3) \delta(q_3-k_2-q_2) \\
    ~&\times &\frac{M_Z^2 ((q_2^2-M^2)+M^2) } {(q_1 ^2-M^2)(q_2 ^2-M^2) (q_3^2-M^2)}   q_{3\mu}q_{3\nu},\nonumber
\end{eqnarray}
\begin{eqnarray}
\label{t31bq}
 ~ & &\frac{1}{M^4} \int d^4q_1 d^4q_2 d^4q_3  (2 \pi)^4 \delta(P-q_1+q_2) \delta(q_1-k_2-q_3) \delta(q_3-k_1-q_2) \\
    ~&\times &\frac{M_Z^2 ((q_1^2-M^2)+M^2) } {(q_1 ^2-M^2)(q_2 ^2-M^2) (q_3^2-M^2)}   q_{3\mu}q_{3\nu}.\nonumber
\end{eqnarray}

But to get a clear and definite cancellation of the quadratic divergence, from the above three equations we still have to again take out the logarithmic divergent ($\frac{M_Z^2}{M^2}$)  terms to be considered in next sections (which is indeed necessary to cancel divergence there, that also guarantees this derivation):
\begin{equation}
\label{the2}
\frac{M_Z^2}{M^2} \int d^4q_1 d^4q_2   (2 \pi)^4  \delta(P-q_1+q_2) \delta(k_1+k_2-q_1+q_2) \frac{- g_{\mu \nu} } {(q_1 ^2-M^2)  (q_2^2-M^2)},
\end{equation}
and
\begin{eqnarray}
\label{the3}
& &\frac{M_Z^2}{M^2} [\int d^4q_1 d^4q_2 d^4q_3  (2 \pi)^4 \delta(P-q_1+q_2) \delta(q_1-k_1-q_3) \delta(q_3-k_2-q_2) \\
\label{the4}
& +& \int d^4q_1 d^4q_2 d^4q_3  (2 \pi)^4 \delta(P-q_1+q_2) \delta(q_1-k_2-q_3) \delta(q_3-k_1-q_2)] \\
& \times & \frac{ q_{3\mu} q_{3\nu} } {(q_1 ^2-M^2) (q_2 ^2-M^2) (q_3^2-M^2)} . \nonumber
\end{eqnarray}

So the quadratic divergence from Eq. (\ref{tcq}) is
\begin{equation}
 \label{taqr}
      T_{Aq}= \frac{M_Z^2}{M^4} (2 \pi)^4 \delta(P-k_1-k_2) \frac{- g_{\mu \nu}}{2} (\int d^4q_1 \frac{1 } {(q_1 ^2-M^2)}+ d^4q_2  \frac{1 } {(q_2 ^2-M^2)}),
\end{equation}
after cancelling the similar factor in the numerator and denominator, and integrating the variable only appearing in the $\delta$  function.

   For the quadratics in  Eqs. (\ref{t11bq}) and (\ref{t31bq}), respectively,
   after cancelling the similar factor in the numerator and denominator, and integrating the variable only appearing in the $\delta$  function,
   they become,
   \begin{equation}
\label{t11bqr}
 T_{11Bq}= \frac{M_Z^2}{M^4} \int d^4q_1  d^4q_3  (2 \pi)^4 \delta(P-q_1+q_3-k_2) \delta(q_1-k_1-q_3)  \frac{ q_{3\mu}q_{3\nu}  } {(q_1 ^2-M^2) (q_3^2-M^2)},
\end{equation}
\begin{equation}
\label{t31bqr}
 T_{31Bq}= \frac{M_Z^2}{M^4} \int  d^4q_2 d^4q_3  (2 \pi)^4 \delta(P+q_2-q_3-k_2)  \delta(q_3-k_1-q_2) \frac{  q_{3\mu}q_{3\nu}  } {(q_2 ^2-M^2) (q_3^2-M^2)}  .
\end{equation}
In the above derivations, we take  many 'petty' steps to extract various terms which is finite and definite or   divergent only logarithmically, and at last arrive at the above three quadratic terms to sum and  cancel. The reason is just that we eschew any shift of the integrated variable in the  integral of high divergence order. The subtle need of them (except the above U(1) invariant $M_Z^2/M^4$ term) to cancel divergence and to get U(1) invariant form is of course a nontrivial guarantee on the derivation.

 The above three terms sum to be   zero, because of the   quadratic surface term (to be zero). Here we come to investigate: 
Similar as  the relation of  logarithmic tensor integral (here we only write the integrand)
\begin{equation}
\label{lgd}
\partial_{(l)}^\mu \frac{l^\nu}{(l^2-\Delta)^2}=\frac{g^{\mu \nu}}{(l^2-\Delta)^2}-\frac{4 l^\mu l^\nu}{(l^2-\Delta)^3}=\frac{l^2 g^{\mu \nu}- 4 l^\mu l^\nu}{(l^2-\Delta)^3}-\frac{\Delta g^{\mu \nu}}{(l^2-\Delta)^3},
\end{equation}
we have the quadratic one
\begin{equation}
\label{qud}
\partial_{(l)}^\mu \frac{l^\nu}{(l^2-\Delta)}=\frac{g^{\mu \nu}}{(l^2-\Delta)}-\frac{2 l^\mu l^\nu}{(l^2-\Delta)^2}=\frac{l^2 g^{\mu \nu}- 2 l^\mu l^\nu}{(l^2-\Delta)^2}-\frac{\Delta g^{\mu \nu}}{(l^2-\Delta)^2}.
\end{equation}
When we take the surface term to be integrated zero, we obtain the tensor reduction formula we need. Why  the surface term can be zero considering the  boundary condition at infinity inherent of the phase space free Feynman propagator \cite{Li:2017hnv},
and  the indication that this momentum space divergence as sensitive probe on local property of space time et vice verse, are to be investigated in the discussion section.
This quadratic form is also appear in QED photon self energy. It is easy for the above formula to give the $U_{EM}(1)$ gauge invariant form (W.I.), proportional only to a log pole which only affect the residue of the photon propagator and is absorbed by the coupling constant renormalization  \cite{Bao:2021byx}. 


By employing Eq. (\ref{qud}),   and introducing a dummy variable $q_3$ and a $\delta$ function, the first term and  second term  in Eq. (\ref{taqr}) can be written as
\begin{equation}
\label{taqr1}
T_{Aq1}= \frac{M_Z^2}{M^4} \int d^4q_1  d^4q_3  (2 \pi)^4 \delta(P-q_1+q_3-k_2) \delta(q_1-k_1-q_3) \frac{- q_{1\mu}q_{1\nu}  } {(q_1 ^2-M^2)^2},
\end{equation}
\begin{equation}
\label{taqr2}
T_{Aq2}= \frac{M_Z^2}{M^4} \int d^4q_2  d^4q_3  (2 \pi)^4 \delta(P+q_2-q_3-k_2) \delta(q_3-k_1-q_2) \frac{- q_{2\mu}q_{2\nu}  } {(q_2 ^2-M^2)^2}.
\end{equation}
We will explicitly calculate that   they 
correspondingly cancel Eq. (\ref{t11bqr}) and Eq. (\ref{t31bqr}), respectively. The integrand of  (\ref{t11bqr})+(\ref{taqr1}) is 
$$\frac{q_{3\mu}q_{3\nu}(q_1^2-q_3^2)}{(q_1^2-M^2)^2(q_3^2-M^2)}+\frac{1}{2}\partial_\mu \frac{k_1\nu}{q_1^2-M^2}
=\frac{q_{3\mu}q_{3\nu} 2 q_1\cdot k_1}{(q_1^2-M^2)^2(q_3^2-M^2)},$$ with the    2nd term  a surface term and  zero after integration. 
This 2nd term  is independent from $q_3$ which can be integrated with one of the $\delta$ functions, leaving full differential hence a  surface term.   In the above equation and in the following the corresponding momentum  of the derivative  will not be written explicitly. 
For the feasibility to go ahead to show they are all surface terms, we rewrite the denominator with  Feynman Parameterization:
$$  \int dx 2x  \frac{ q_{1\mu}(q_{1\nu}-k_{1\nu}) 2 q_1\cdot k_1}{((q_1+(x-1) k_1)^2- M^2) ^3}. $$  Here we have employed   $ k_1^2=0$ and $k_{1\mu}=0$. The above  equals to the following
$$  \int dx \frac{-1}{2}x (\partial_{\mu} \frac{(q_{1\nu}-k_{1\nu}) 2 q_1\cdot k_1}{((q_1+(x-1) k_1)^2- M^2) ^2} + \frac{(g_{\mu\nu}) 2 q_1\cdot k_1}{((q_1+(x-1) k_1)^2- M^2) ^2} + \frac{(q_{1\nu}-k_{1\nu}) 2  k_{1\mu}}{((q_1+(x-1) k_1)^2- M^2) ^2}). $$
The first one is surface term (=0 after integration). The third term is zero since $k_{1\mu}=0$.
The second term  is again  surface term: $$- \partial_\xi \frac{(g_{\mu\nu})  k_1^\xi}{((q_1+(x-1) k_1)^2- M^2) }$$ considering $ k_1^2=0$. It is again zero after integration.
 
The integrand of  (\ref{t31bqr})+ (\ref{taqr2}) is just the similar and the integral =0.

Here we emphasize again this petty derivation  just to show the importance of the surface term and the possibility to eschew any shift for high order divergence.

Now we see that the total result on the terms proportional to $M_Z^2/M^4$ is  finite, non-zero, and U(1) gauge invariant. 
And the $M_Z^2/M^2$ terms  left after all the above cancellation are to be shown necessary for the following cancellation. 


   %
From the above,  we learn that the calculation in the way introduced in this paper, especially not to integrate the $\delta$ functions before have to  and ia allowed to (once a variable only appearing in the delta function, it is a dummy and the integrand  is independent from it),
provides the exact definition of the Feynman diagram.
%
%
%
%
%
One may suspect that for the most general case of the calculation of the Feynman diagrams, the proper way of
setting the independent
integrated  variables at beginning as done by GWW \cite{Gastmans:2011ks},  may not be available. So calculation without integration on  the $\delta$ functions until having to is a more proper or maybe necessary way of the employment of the Feynman rules.
 %
%
As for the surface term, it is consistent with the na\"ive symmetric integration for convergent integral.  The  surface integral formula is also consistent with a D dimension regularization calculation (which is always considered as convergent), and share the similar spirit of the 'IBP'.   
Though it is unclear that All  calculations including the above and other similar stuffs done in D dimension can always get the same result, some  statements on the 'discontinuity' between D and four dimensions like \cite{dr-4}, is  incorrect mainly coming from the incorrect employment of the symmetric integration  for the case of  logarithmic divergence because of incorrect treating the surface term, as in the  Jauch and Rohrlich book. Similar case  is for the momentum shifting in  linear divergence, which has been pointed out in discussion section of \cite{Bao:2021byx}.

 \subsection{}
 Now here we investigate the $M^{-2}$ terms ($T_{113}$ (\ref{t113}), $T_{313}$ (\ref{t313}), $T_{11B1}$ (\ref{t11b1}),  $T_{31B1}$ (\ref{t31b1}), and Eqs. (\ref{the2}), (\ref{the3}), (\ref{the4}) included).


Besides the above listed $M^{-2}$ terms in bracket, we need to investigate the following:  From $T_1$
\begin{eqnarray}
\label{t12}
T_{12} &= &  \frac{-1}{M^2} \int d^4q_1 d^4q_2 d^4q_3 (2 \pi)^4 \delta(P-q_1+q_2) \delta(q_1-k_1-q_3)\delta(q_3-k_2-q_2)   \\
~& \times & g_\alpha~^{\beta}  g ^{\rho \sigma} q_2^\alpha q_2 ^\gamma
 \frac{V_{\beta \mu \rho} (q_1,  -k_1, -q_3) ~ V_{\sigma \nu \gamma}(q_3, -k_2, -q_2)} {(q_1 ^2-M^2)(q_3 ^2-M^2)(q_2 ^2-M^2)}, \nonumber
\end{eqnarray}
\begin{eqnarray}
\label{t13}
 T_{13} &= &  \frac{-1}{M^2} \int d^4q_1 d^4q_2 d^4q_3 (2 \pi)^4 \delta(P-q_1+q_2) \delta(q_1-k_1-q_3)\delta(q_3-k_2-q_2)   \\
~& \times & g_\alpha~^{\beta}  q_3 ^\rho q_3 ^\sigma g^{\alpha \gamma}
 \frac{V_{\beta \mu \rho} (q_1,  -k_1, -q_3) ~ V_{\sigma \nu \gamma}(q_3, -k_2, -q_2)} {(q_1 ^2-M^2)(q_3 ^2-M^2)(q_2 ^2-M^2)}, \nonumber
\end{eqnarray}
\begin{eqnarray}
\label{t14}
 T_{14} &= &  \frac{-1}{M^2} \int d^4q_1 d^4q_2 d^4q_3 (2 \pi)^4 \delta(P-q_1+q_2) \delta(q_1-k_1-q_3)\delta(q_3-k_2-q_2)   \\
~& \times & q_{1\alpha} q_1 ^{\beta}g ^{\rho \sigma}g^{\alpha \gamma}
 \frac{V_{\beta \mu \rho} (q_1,  -k_1, -q_3) ~ V_{\sigma \nu \gamma}(q_3, -k_2, -q_2)} {(q_1 ^2-M^2)(q_3 ^2-M^2)(q_2 ^2-M^2)}. \nonumber
\end{eqnarray}
 $T_{12}$ and $T_{14}$ both give $q_3^2-M^2$ term in numerator when applying the Ward identity directly, which  could reduce the corresponding factor in
denominator and combine with $T_{22+23}$,
\begin{eqnarray}
\label{t22e3}
 T_{22+23} &= &  \frac{1}{M^2} \int d^4q_1 d^4q_2  (2 \pi)^4 \delta(P-q_1+q_2) \delta(q_1-q_2-k_1-k_2)   \\
~& \times &  \frac{2q_1^2 g_{\mu \nu} + 2q_2^2 g_{\mu \nu}-2 q_{1 \mu} q_{1 \nu}-2 q_{2 \mu} q_{2 \nu}} {(q_1 ^2-M^2)(q_2 ^2-M^2)}.  \nonumber
\end{eqnarray}
$T_{12}=T_{121}+T_{122}+T_{123}+T_{124}$:
\begin{eqnarray}
\label{t121}
T_{121} &= &  \frac{-1}{M^2} \int d^4q_1 d^4q_2 d^4q_3 (2 \pi)^4 \delta(P-q_1+q_2) \delta(q_1-k_1-q_3)\delta(q_3-k_2-q_2)   \\
~& \times &  q_2^\beta
 \frac{V_{\beta \mu \nu} (q_1,  -k_1, -q_3)} {(q_1 ^2-M^2)(q_2 ^2-M^2)}; \nonumber
\end{eqnarray}
\begin{eqnarray}
T_{122} &= &  \frac{-1}{M^2} \int d^4q_1 d^4q_2 d^4q_3 (2 \pi)^4 \delta(P-q_1+q_2) \delta(q_1-k_1-q_3)\delta(q_3-k_2-q_2)   \\
~& \times &  q_2^\beta
 \frac{V_{\beta \mu \sigma} (q_1,  -k_1, -q_3) (-q_3 ^\sigma) q_{3\nu}} {(q_1 ^2-M^2)(q_3 ^2-M^2)(q_2 ^2-M^2)}; \nonumber 
\end{eqnarray}
\begin{eqnarray}
T_{123} &= &  \frac{-1}{M^2} \int d^4q_1 d^4q_2 d^4q_3 (2 \pi)^4 \delta(P-q_1+q_2) \delta(q_1-k_1-q_3)\delta(q_3-k_2-q_2)   \\
~& \times &  q_2^\beta
 \frac{V_{\beta \mu \nu} (q_1,  -k_1, -q_3) M^2} {(q_1 ^2-M^2)(q_3 ^2-M^2)(q_2 ^2-M^2)} \nonumber 
\end{eqnarray}
is $M^0$ term;
\begin{eqnarray}
T_{124} &= &  \frac{-1}{M^2} \int d^4q_1 d^4q_2 d^4q_3 (2 \pi)^4 \delta(P-q_1+q_2) \delta(q_1-k_1-q_3)\delta(q_3-k_2-q_2)   \\
~& \times &  q_2^\beta
 \frac{V_{\beta \mu \nu} (q_1,  -k_1, -q_3) (-M_Z^2)} {(q_1 ^2-M^2)(q_3 ^2-M^2)(q_2 ^2-M^2)}. \nonumber 
\end{eqnarray}
 $T_{14}=T_{141}+T_{142}+T_{143}$:
 \begin{eqnarray}
 \label{t141}
 T_{141} &= &  \frac{-1}{M^2} \int d^4q_1 d^4q_2 d^4q_3 (2 \pi)^4 \delta(P-q_1+q_2) \delta(q_1-k_1-q_3)\delta(q_3-k_2-q_2)   \\
~& \times &
 \frac{V_{\mu \nu \gamma} (q_3,  -k_2, -q_2)} {(q_1 ^2-M^2)(q_2 ^2-M^2)}q_1^\gamma; \nonumber
\end{eqnarray}
 \begin{eqnarray}
T_{142} &= &  \frac{1}{M^2} \int d^4q_1 d^4q_2 d^4q_3 (2 \pi)^4 \delta(P-q_1+q_2) \delta(q_1-k_1-q_3)\delta(q_3-k_2-q_2)   \\
~& \times &
 \frac{q_{3\mu} q_3 ^\sigma V_{\sigma \nu \gamma} (q_3,  -k_2, -q_2) } {(q_1 ^2-M^2)(q_3 ^2-M^2)(q_2 ^2-M^2)} q_1^\gamma; \nonumber 
\end{eqnarray}
 \begin{eqnarray}
T_{143} &= &  \frac{-1}{M^2} \int d^4q_1 d^4q_2 d^4q_3 (2 \pi)^4 \delta(P-q_1+q_2) \delta(q_1-k_1-q_3)\delta(q_3-k_2-q_2)   \\
~& \times &
 \frac{M^2 V_{\mu \nu \gamma} (q_3,  -k_2, -q_2) } {(q_1 ^2-M^2)(q_3 ^2-M^2)(q_2 ^2-M^2)} q_1^\gamma \nonumber 
\end{eqnarray}
is $M^0$ term.

This discussion also applies to $T_3$, so the following is to investigate $T_{121}+T_{141}+T_{321}+T_{341}+T_{22+23}$,  and we use
$q_2=q_1-P$,  so can use W.I. again, while the extra term    $-P^\beta
 V_{\beta \mu \nu} (q_1,  -k_1, -q_3)$  will combine with the corresponding term from $T_{141}$.
We use $q_1=q_2+P$, and the extra term is $V_{\mu \nu \gamma} (q_3,  -k_2, -q_2)P^\gamma $ there.
 We found that the extra $M_Z^2$ term from the W.I. and from $(k_1+k_2)^2$  just cancelled, so all the other similar as two photon case \cite{Li:2017hnv}, i.e.,
  %
   %
$$T_{121}+T_{141}+T_{321}+T_{341}+T_{22+23}=0.$$

~~~~~~


~~~~~~~~~~~~~~

~~~~~~~~~~~~~~~~~~~~~~~~~

Now the remaining $M^{-2}$ terms are all from $T_1$ and $T_3$, {\it as well as the remained $M_Z^2$ terms from the above subsection.} 
   %
 As the case of the two photons, cancellation for the linear divergence will be taken by the summation of the corresponding terms from   $T_1$ and $T_3$ respectively.
Those directly from $T_{1}$  are \begin{equation} \label{m220} 
 T_{113}+T_{13}+T_{122}+T_{124}+T_{142}, \end{equation} which  must  be considered together with $M_Z^2$ terms   half of Eq. (\ref{the2}),  then $T_{11B1}$ (\ref{t11b1}) and Eq. (\ref{the3}).
 It is easy to find that the terms without the $M_Z^2$ factor are quite similar as two photon case, but extra terms from $k_2^2=M_Z^2$ are subtle.
 They emerge from various equations, and cancel in various ways leading to the U(1) invariant final result, which is a manifest of the self-consistent of the standard model.

So we separate (\ref{m220}) as two parts: A, those  explicitly without $M^2_Z$ factor in the beginning; B, those with. Then we calculate A, to see which canceled, which giving extra $M_Z^2$ terms to be canceled
with extra terms from last section, or to be arranged with part B---together with all extra terms from last subsection, leading to the final U(1) invariant results.

We here again  apply the W.I. and combine them together to obtain a simple form of part A:
\begin{eqnarray}
~ & ~ & (T_{113}+T_{13}+T_{122}+T_{142} )_A \\
~& = & \frac{1}{M^2} \int d^4q_1 d^4q_2 d^4q_3 \frac{  (2 \pi)^4 \delta(P-q_1+q_2) \delta(q_1-k_1-q_3)\delta(q_3-k_2-q_2) }  {(q_1 ^2-M^2)(q_3 ^2-M^2)(q_2 ^2-M^2)} \nonumber \\
 ~& \times & (2 q_1^2 q_{2\mu} q_{2\nu} + 2 q_2^2 q_{1\mu} q_{1\nu} -4 q_1\cdot q_2 q_{1\mu} q_{2\nu} + q_1\cdot q_2 q_3^2 g_{\mu \nu}-q_1^2 q_2 ^2 g_{\mu \nu}). \nonumber
\end{eqnarray}
It looks as quadratic, but easy to see  in fact to the most linear, quite similar  as the two photon case,
since

$2 q_1\cdot q_2=-(q_1-q_2)^2+q_1^2+q_2^2, $  then
$ 2 q_1^2 q_{2\mu} q_{2\nu} + 2 q_2^2 q_{1\mu} q_{1\nu} -4 q_1\cdot q_2 q_{1\mu} q_{2\nu}$
equals to
 $$2 q_1^2(q_{2\mu}-q_{1\mu})q_{2\nu} +2 q_2^2 q_{1\mu}(q_{1\nu}-q_{2\nu})
+2 (q_1-q_2)^2q_{1\mu}q_{2 \nu}=2 q_1^2 (-k_{2\mu})q_{2\nu} +2 q_2^2 q_{1\mu}k_{1\nu}
+2 (k_1+k_2)^2 q_{1\mu}q_{2 \nu},$$

and
$q_1\cdot q_2 q_3^2 g_{\mu \nu}-q_1^2 q_2 ^2 g_{\mu \nu}=-\frac{(k_1+k_2)^2}{2}q_3^2 g_{\mu \nu}+\frac{q_1^2+q_2^2}{2}q_3^3 g_{\mu \nu}-q_1^2 q_2 ^2 g_{\mu \nu}.$

However, $$\frac{q_1^2+q_2^2}{2}q_3^3 g_{\mu \nu}-q_1^2 q_2 ^2 g_{\mu \nu}= (\frac{q_1^2}{2} (q_3+q_2)\cdot k_2 -\frac{q_2^2}{2} (q_3+q_1)\cdot k_1)g_{\mu \nu}$$ hence is also linear ($q_3$ and $q_2$ can not combine since $k_2^2=M^2_Z\neq 0$).


Now we write
$(T_{113}+T_{13}+T_{122}+T_{142})_A$  
 as the summation of two parts:
 \begin{eqnarray}
 \label{line}
  &  & \frac{1}{M^2} \int d^4q_1 d^4q_2 d^4q_3 \frac{  (2 \pi)^4 \delta(P-q_1+q_2) \delta(q_1-k_1-q_3)\delta(q_3-k_2-q_2) }  {(q_1 ^2-M^2)(q_3 ^2-M^2)(q_2 ^2-M^2)} \nonumber \\
 ~& \times & 2 q_1^2 (-k_{2\mu})q_{2\nu} +2 q_2^2 q_{1\mu}k_{1\nu}+ (\frac{q_1^2}{2} (q_3+q_2)\cdot k_2 -\frac{q_2^2}{2} (q_3+q_1)\cdot k_1)g_{\mu \nu}, 
\end{eqnarray}
\begin{eqnarray}
\label{loog}
  &  & \frac{1}{M^2} \int d^4q_1 d^4q_2 d^4q_3 \frac{  (2 \pi)^4 \delta(P-q_1+q_2) \delta(q_1-k_1-q_3)\delta(q_3-k_2-q_2) }  {(q_1 ^2-M^2)(q_3 ^2-M^2)(q_2 ^2-M^2)} \nonumber \\
 ~& \times & 2 (k_1+k_2)^2 q_{1\mu}q_{2 \nu}-\frac{(k_1+k_2)^2}{2}q_3^2 g_{\mu \nu}, 
\end{eqnarray}
i.e., the linear and logarithmic divergent terms respectively.
Then the above linear term, after further taking out   logarithmic  and  finite terms from it,
 should  combine with the corresponding term from $T_3$, then is deduced   to get  as two terms,  one is logarithmic divergent, the other is finite.

 Some details are:
 \begin{eqnarray}
  2 q_1^2 (-k_{2\mu})q_{2\nu}&=&-2(q_3^2-M^2)k_{2\mu}q_{2\nu}-4q_3\cdot k_1 k_{2\mu}q_{2\nu}-2M^2k_{2\mu}q_{2\nu}   \\
 2 q_2^2 q_{1\mu}k_{1\nu}&=& 2 (q_3^2-M^2) q_{1\mu}k_{1\nu} -4 q_3\cdot k_2 q_{1\mu}k_{1\nu} +2 M^2  q_{1\mu}k_{1\nu}+2 M_Z^2  q_{1\mu}k_{1\nu}  \\
 \frac{q_1^2}{2} (q_3+q_2)\cdot k_2 g_{\mu \nu}&=&\frac{(q_3^2-M^2)}{2} (q_3+q_2)\cdot k_2  g_{\mu \nu}
 + q_3\cdot   k_1(q_3+q_2)\cdot k_2  g_{\mu \nu} \nonumber \\
 & +& \frac{M^2}{2}(q_3+q_2)\cdot k_2  g_{\mu \nu}  \\
  -\frac{q_2^2}{2} (q_3+q_1)\cdot k_1g_{\mu \nu}&=&\frac{-(q_3^2-M^2)}{2} (q_3+q_1)\cdot k_1  g_{\mu \nu}
 + q_3\cdot   k_2(q_3+q_1)\cdot k_1 g_{\mu \nu} \nonumber \\
 & -& \frac{M^2}{2}(q_3+q_1)\cdot k_1 g_{\mu \nu}   - \frac{M_Z^2}{2}(q_3+q_1)\cdot k_1 g_{\mu \nu}
  \end{eqnarray}
  First of all we address that the $M_Z^2$ terms (some are not explicitly written in the above) are all finite terms and are to be calculated later. 
So it is the  following linear term (with $(q_3^2-M^2)$ factor reduced with the common one in denominator, and $(q_3+q_2)$ written as $(2q_2+k_2)$, $(q_3+q_1)$ written as $(2q_1-k_1)$, then $q_3$ integrated)
\begin{eqnarray}
  &~& \frac{1}{M^2} \int d^4q_1 d^4q_2  \frac{  (2 \pi)^4 \delta(P-q_1+q_2) \delta(q_1-q_2-k_1-k_2)}  {(q_1 ^2-M^2)(q_2 ^2-M^2)}
 ( -2k_{2\mu}q_{2\nu} +2  q_{1\mu}k_{1\nu} \nonumber \\
& +& \frac{1}{2}((2q_2+k_2)\cdot k_2 -(2q_1-k_1) \cdot k_1)) g_{\mu \nu})
 \end{eqnarray}
 to be combined with that from $T_3$:
\begin{eqnarray}
   &~& \frac{1}{M^2} \int d^4q_1 d^4q_2  \frac{  (2 \pi)^4 \delta(P-q_1+q_2) \delta(q_1-q_2-k_1-k_2)}  {(q_1 ^2-M^2)(q_2 ^2-M^2)}
 ( -2k_{1\nu}q_{2\mu} +2  q_{1\nu}k_{2\mu} \nonumber \\
  &+& \frac{1}{2}((2q_2+k_1)\cdot k_1 -(2q_1-k_2) \cdot k_2)) g_{\mu \nu}),
 \end{eqnarray}
 {\it  and summed with Eq. (\ref{the2}) (!)}
then  deduces to logarithmic term. {\it Half} of their summation is then:
\begin{equation}
  \frac{1}{M^2} \int d^4q_1 d^4q_2  \frac{  (2 \pi)^4 \delta(P-q_1+q_2) \delta(q_1-q_2-k_1-k_2)}  {(q_1 ^2-M^2)(q_2 ^2-M^2)}
( 2  k_{2\mu}k_{1\nu} - \frac{(k_1+ k_2)^2}{2}  g_{\mu \nu})
\end{equation}
This term can again be separated into a logarithmic term and a finite term,
\begin{eqnarray}
 \frac{1}{M^2} \int d^4q_1 d^4q_2 d^4q_3 \frac{  (2 \pi)^4 \delta(P-q_1+q_2) \delta(q_1-k_1-q_3)\delta(q_3-k_2-q_2) }  {(q_1 ^2-M^2)(q_3 ^2-M^2)(q_2 ^2-M^2)}
q_3^2 ( 2  k_{2\mu}k_{1\nu} - \frac{(k_1 + k_2)^2}{2} g_{\mu \nu}) \nonumber \\
+\frac{1}{M^2} \int d^4q_1 d^4q_2 d^4q_3 \frac{  (2 \pi)^4 \delta(P-q_1+q_2) \delta(q_1-k_1-q_3)\delta(q_3-k_2-q_2) }  {(q_1 ^2-M^2)(q_3 ^2-M^2)(q_2 ^2-M^2)}
(-M^2) ( 2  k_{2\mu}k_{1\nu} - \frac{(k_1 + k_2)^2}{2} g_{\mu \nu})  \nonumber
 \end{eqnarray}
 Hence {\it effectively}
 $(T_{113}+T_{13}+T_{122}+T_{142})_A +Eq. (\ref{the2})/2 =T1LG_A+T1LG_B+T1F_A+T1F_B$, with
 \begin{eqnarray}
 \label{t1lg}
 T1LG_A = \frac{1}{M^2} \int d^4q_1 d^4q_2 d^4q_3 \frac{  (2 \pi)^4 \delta(P-q_1+q_2) \delta(q_1-k_1-q_3)\delta(q_3-k_2-q_2) }  {(q_1 ^2-M^2)(q_3 ^2-M^2)(q_2 ^2-M^2)} \nonumber\\
 \times  [(- 2 q_3^2 k_1\cdot k_2 + 4 k_1 \cdot q_3 k_2 \cdot q_3) g_{\mu \nu}-4 k_1 \cdot q_3 k_{2 \mu} q_{3 \nu}-4 k_2 \cdot q_3 q_{3 \mu} k_{1 \nu}
 +2 q_3^2 k_{2\mu} k_{1\nu} +4 k_1 \cdot k_2 q_{3\mu} q_{3 \nu}]\nonumber \\
 ~
 \end{eqnarray}
  %
  \begin{eqnarray}
 T1F_A&=&  \int d^4q_1 d^4q_2 d^4q_3 \frac{  (2 \pi)^4 \delta(P-q_1+q_2) \delta(q_1-k_1-q_3)\delta(q_3-k_2-q_2) }  {(q_1 ^2-M^2)(q_3 ^2-M^2)(q_2 ^2-M^2)}\\
 &\times & [2 q_{1\mu} k_{1\nu}-2k_{2\mu}q_{2\nu}-2k_{2\mu}k_{1\nu} +\frac{1}{2}((q_3+q_2)\cdot k_2-(q_3+q_1) \cdot k_1+(k_1+ k_2)^2  )g_{\mu \nu}], \nonumber
 \end{eqnarray} ($M^0$ term to be combined into the final result), and  U(1) invariant finite
  \begin{eqnarray}
 T1F_B&=& \frac{M_Z^2}{M^2} \int d^4q_1 d^4q_2 d^4q_3  (2 \pi)^4 \delta(P-q_1+q_2) \delta(q_1-k_1-q_3)\delta(q_3-k_2-q_2)\\
 &\times & \frac{2 q_{1\mu} k_{1\nu} -2 q_1 \cdot k_1g_{\mu \nu}  }  {(q_1 ^2-M^2)(q_3 ^2-M^2)(q_2 ^2-M^2)}. \nonumber
 \end{eqnarray}

But the $T1LG_B$ term is the logarithmic terms with $M_Z^2$ factor,
 summing the $M_Z^2$ terms $(T_{113}+T_{13}+T_{124}+T_{142})_B$,   {\it    Eq. (\ref{the3}) , and $T_{11B1}$ (\ref{t11b1})}.  
The result is
\begin{eqnarray}
 T1F_C&= &\frac{M_Z^2}{M^2} \int d^4q_1 d^4q_2 d^4q_3  (2 \pi)^4 \delta(P-q_1+q_2) \delta(q_1-k_1-q_3)\delta(q_3-k_2-q_2) \\\nonumber
 &\times &   \frac{2 k_1\cdot k_2 g_{\mu\nu}-2 k_{2\mu}k_{1\nu} } {(q_1 ^2-M^2)(q_3 ^2-M^2)(q_2 ^2-M^2)}.
 \end{eqnarray}
 The $T_3$ is in fact effectively discussed, and just make the above doubled.
We see that all the above are finite and U(1) invariant ($T1F_A$ is to be combined in the following), and {\it  all the remaining ones in last subsection  has been cancelled and combined}.
This procedure is subtle, interesting and self-consistent (due to standard model).
From this subsection we obtain U(1) EM gauge invariant terms proportional to $1/M^2$, $T1LG_A$ (the logarithmic cancelled so finite) as the two photon case, we mark as R2. We also obtain the extra finite U(1) EM invariant terms
proportional to $M_Z^2/M^2$, $T1F_B$ and $T1F_C$, we mark as R3, R4.  These are also to be considered for the investigation of the gauge invariant w.r.t. $R_\xi$ gauge result.

\subsection{} Now the $M^0$ terms.

The third part of $T_{12},~ T_{14}$, i.e., $T_{123}, ~ T_{143}$, as well as those corresponding ones from $T_3$, are $M^0$ terms and to be investigated here
together with the corresponding terms from $T_2$   
  (Pay attention that $T_2$ lack of a overall minus sign) and the other remaining $M^0$ terms form $T_1$ and $T_3$:
  %
  %
\begin{eqnarray}
T_{15}&=&\int d^4q_1 d^4q_2 d^4q_3 (2 \pi)^4 \delta(P-q_1+q_2) \delta(q_1-k_1-q_3)\delta(q_3-k_2-q_2)  \\
~&\times & g_\alpha~^{\beta} g ^{\rho \sigma} g^{\alpha \gamma}
 \frac{V_{\beta \mu \rho} (q_1,  -k_1, -q_3) ~ V_{\sigma \nu \gamma}(q_3, -k_2, -q_2)} {(q_1 ^2-M^2)(q_3 ^2-M^2)(q_2 ^2-M^2)}, \nonumber
\end{eqnarray}
and $T_{35}$ is not explicitly written here.
\begin{eqnarray}
 T_{24} &= & - \int d^4q_1 d^4q_2  (2 \pi)^4 \delta(P-q_1+q_2) \delta(q_1-q_2-k_1-k_2)   \\
~& \times & g_\alpha~^{\beta}  g^{\alpha \gamma}
 \frac{2g_{\mu \nu} g_{\beta \gamma}-g_{\mu \beta} g_{\nu \gamma} -g_{\mu \gamma} g_{\nu \beta}} {(q_1 ^2-M^2)(q_2 ^2-M^2)}. \nonumber
\end{eqnarray}

These four terms (half of $T_{24}$) summed  still give  terms as logarithmic
\begin{eqnarray}
~&~& \int d^4q_1 d^4q_2 d^4q_3  (2 \pi)^4 \delta(P-q_1+q_2) \delta(q_1-k_1-q_3)\delta(q_3-k_2-q_2)\nonumber \\
~&\times & \frac{(-3 q_3^2 g_{\mu \nu}+12 q_{3\mu} q_{3\nu})} {(q_1 ^2-M^2)(q_3 ^2-M^2)(q_2 ^2-M^2)},
\end{eqnarray}
but   all the divergences cancelled, and the total summation of the terms, half of $T_{24}$, $T_{15}$, $T_{123}, ~ T_{143}$, as well as
$T1F_A$, give the exactly similar form of  Eq. (36) of \cite{Li:2017hnv} ($H \to \gamma \gamma$ process), i.e.,
\begin{eqnarray}
\label{r1}
~&~& \int d^4q_1 d^4q_2 d^4q_3  (2 \pi)^4 \delta(P-q_1+q_2) \delta(q_1-k_1-q_3)\delta(q_3-k_2-q_2)\nonumber \\
~&\times & \frac{-6 k_1\cdot k_2 g_{\mu \nu} + 6 k_{2\mu}k_{1\nu} + 3M^2 g_{\mu \nu} + (-3 q_3^2 g_{\mu \nu}+12 q_{3\mu} q_{3\nu})} {(q_1 ^2-M^2)(q_3 ^2-M^2)(q_2 ^2-M^2)}.
\end{eqnarray}
But pay attention since the denominator different because of $k_2^2=M_Z^2\neq 0$, this is U(1) invariant  just after some of the terms integrated to be zero.  However, same as two photon case, the Dyson subtraction is
not needed since the surface term formula for log divergence.
We mark this term as R5. 

In this paper we do not intend to produce the final results. Investigation on $SU(2) \times U(1)$ gauge invariance and combining fermion loops are to be done in the following paper. In this paper we just give the original form we obtained above, R1+R2+R3+R4+R5.






    %
  %

  \section{Discussion and speculation}
In this paper we eliminate the  paradoxes and ambiguities via the framework of the original Dyson scheme, each propagator momentum integration kept and each $\delta$ function at each vertex not integrated out:
1,  the original loop momentum definite  in    the beginning to write down  the amplitude  in momentum space with Feynman rules;
2,  any momentum shift for divergences worse than logarithmic eschewed;
3,  no regularization or dimension extrapolation.
 %
Some speculation of new physics beyond the standard model can be based on some new symmetry which introduces some correspondence to the known standard model
particles,
and  may simplify the renormalization dramatically via divergence cancellation, especially high order divergence cancelation.
  %
So to properly write down  these divergences and then to properly cancel them become a very crucial point.

To proceed the concrete calculation, the evaluation on the  surface terms \cite{Ferreira:2011cv} for  the   loop momentum going to infinite is very important to eliminate any uncertainty. We refer to the physical boundary condition of the scattering S matrix static problem  to determine the surface term to be zero \cite{Li:2017hnv} which is taken as undetermined in \cite{Ferreira:2011cv}.
In this paper we encounter and determine the  quadratic (\ref{qud}) and logarithmic (\ref{lgd}) ones. In all these problems, the physics in  the Minkowski spacetime free Feynman propagator is very important. The fact that the Feynman propagator is defined on the free particle Hilbert space (constructed by Poincar\'{e} symmetric on shell  Fock state), renders it a special contour integral on the $l^0$ complex plane (it is feasible to use $l^0$ because of the SO(3) symmetry as part of Poincar\'{e} symmetry). So we have the freedom to take the infinite surface with $l^0 \gg |\vec{l}|$. 
Besides the $H \to \gamma \gamma$, $H \to \gamma Z$, the photon and photon-Z self-energy diagram \cite{Bao:2021byx},
these also appear in other places, as for the axial anomaly.  The Dyson scheme and the 'most symmetric  loop momentum' (see footnote 3) are  both in contrary to the Bell-Jackiw claim.  There,  no necessity  referring to the surface term,  but  we can see the surface term determined same as above via the physical boundary condition of the free propagator, gives  the consistency  \cite{Bao:2021byx}.

 However, one must adopt that in the most general consideration which is not restricted in the physical boundary condition of the free propagator in Minkowski momentum space (the scattering  S matrix static problem),   whether the surface term equals to zero depends on the
 concrete condition of the infinite momentum surface.  Hence the  space-time
structure, geometry or topology, correspondingly   are very crucial.
For example, when the divergence of the chiral current   is calculated via the concrete QFT defined on a specified  space-time manifold, the value can  be zero or not. This is in fact  {\it taken as an input} to generate many interesting physics which have been widely studied, e.g., relation  with the  Atiyah-Singer index/Pontryagin class, 't Hoot symmetry breaking, etc.
What lesson we learn here is that for a concrete physical problem,
the Atiyah-Singer index in no way must  equal to nonzero integer, but must be determined by the proper calculation of the QFT defined on the manifold, as well as the physical condition, especially for special limits.
As a classical  original example, the Polyakov solution in Euclidean space-time \cite{Belavin:1975fg}.  Here we only want to point out that besides the index $q \neq 0$  case which gives many interesting speculations so far, the $q=0$  case  also exist and can be considered as special limit of the $q=1$ solution.
One trivial case is  the group element g in the paper independent from the space-time at the large sphere $S_3$, i.e.,  becoming a global gauge. This is more consistent with  confinement than pure gauge.
Another case is  when  $x_4=- it$  can be neglected, index q again can be calculated to be zero. This may show a triviality of the high energy behaviour with the axial current conservation,  sharing similar spirit as the  divergence cancellation in our present paper with  Energy  larger larger than 3-momentum on the infinite momentum surface.

So it is important to clarify the concrete condition for the divergence (un)cancellation, the surface term ($=$ or $\neq$ 0).
The concrete manifold of space-time can make zero nonzero, can make finite nonfinite, can
 make  a renormalizable field theory non renormalizable, or need more operators to close, etc.
Besides others where the anomaly is very important,  considering the 't Hooft way to get charge non-conservation. This can
explain the non-conservation of U(1) number and CP violation of  early universe, and conservation  restoration now, based the difference of the property of the space-time.
In other words, divergence integral (or cancellation, or the concrete value of the surface term) in the calculation of the QFT in momentum space is a sensitive probe on the topology or some other local structure (holes...)  of
the space-time.    




For the study of the cosmology,
  the above discussion on the (ultraviolet) divergence cancellation possibly depending on the concrete (local) structure of spacetime implies interesting speculation.  The 'historical versions'  of  QFT's vary with the  universe  evolution. QFT's  can be defined on various spacetime math structure. In more details,
 different versions of QFT's, defined on different manifolds or more general math structures corresponding to our universe---spacetime---at various special periods can be self -consistent or -inconsistent.
 But in whatsoever cases,
the divergences may cancel to get finite (including zero) prediction or may not and can not give clear prediction, it can always  probe the math structure of the spacetime of the universe. The ultraviolet probes the local, the infrared probes the global, or just manifests the special information of the space time which leads to the inconsistency between the spacetime and QFT.
In the more concrete cases addressed above in the paper, the spacetime in early universe manifold may have singularity, defect, bubble wall or other structure to cause the anomaly to give the CP violation and U(1) charge nonconservation for baryogenesis but these structures of the spacetime manifold now evolute/expand so that locally become Minkowski and no anomaly, no violations but only conservation of the corresponding currents, which is consistent with experiments and has to ask for cancellation in some theoretical  framework without  elimination inherently.  At the same time these early defects or structures can also play the role of  'seed' of the curvature of spacetime (primordial curvature perturbation) via the homogeneous Einstein equations, i.e., this curvature is  not caused by 'matter' (inhomogeneous term in Einstein equations) but is now observed as 'dark matter' when the universe evolute to today.


\section*{Acknowledgments}
The authors greatly thank  Prof. Tai Tsun Wu for stimulating the topic related with the gauge paradox, for encouragement  and   many instructive discussions.



This work is supported in part by  National Natural Science Foundation of China (grant No. 12275157, 11775130, 11635009). 

\appendix
 \section{Mathematics for $H \to \gamma Z$ corresponding to Eqs. (2.5-2.12) of GWW \cite{Gastmans:2011wh}}
(These formulation  will recover to the complemented  GWW formulations which we employed in the above for  calculating the $H \to \gamma \gamma $ process once taking $M_Z=0$.)
\begin{equation}
\label{a1}
k_1^2=0, ~ k_2^2=M_Z^2; \hspace{ 3 cm}  k_{1\mu}=k_{2\nu}=0.
\end{equation}
\begin{equation}
(k_1+k_2)^2=2 k_1 \cdot k_2 +M_Z^2=M_H^2.
\end{equation}
\begin{eqnarray}
V_{\alpha \beta \gamma} (p_1, p_2, p_3)= (p_2-p_3)_{\alpha}g_{\beta \gamma} +(p_3-p_1)_{\beta}g_{\gamma \alpha}+(p_1-p_2)_{\gamma}g_{\alpha \beta}; \\
p_1+p_2+p_3=0~~ (incoming). \nonumber
\end{eqnarray}
\begin{eqnarray}
p_1^{\alpha} V_{\alpha \beta \gamma} (p_1, p_2, p_3)=(p_3^2g_{\beta \gamma}-p_{3 \beta}p_{3 \gamma})-(p_2^2g_{\beta \gamma}-p_{2 \beta}p_{2 \gamma})\\
V_{\alpha \beta \gamma} (p_1, p_2, p_3) p_3 ^\gamma =-(p_1^2 g_{\alpha \beta}-p_{1\alpha} p_{1\beta}) +(p_2^2 g_{\alpha \beta}-p_{2\alpha} p_{2\beta}) \nonumber
\end{eqnarray}
\begin{eqnarray}
p_1^{\alpha} V_{\alpha \mu \gamma} (p_1, -k_1, p_3)=p_3^2g_{\mu \gamma}-p_{3 \mu}p_{3 \gamma}\\
V_{\alpha \mu \gamma} (p_1, -k_1, p_3) p_3 ^\gamma =-(p_1^2 g_{\alpha \mu}-p_{1\alpha} p_{1\mu})\nonumber \\
p_1^{\alpha} V_{\alpha \nu \gamma} (p_1, -k_2, p_3)=p_3^2g_{\nu \gamma}-p_{3 \nu}p_{3 \gamma}-M_Z^2 g_{\nu \gamma}\\
V_{\alpha \nu \gamma} (p_1, -k_2, p_3) p_3 ^\gamma =-(p_1^2 g_{\alpha \nu}-p_{1\alpha} p_{1\nu})+M_Z^2 g_{\alpha \nu}\nonumber
\end{eqnarray}
\begin{eqnarray}
\label{a7}
p_1^{\alpha} V_{\alpha \mu \gamma} (p_1, -k_1, p_3)=(p_3^2-M^2)g_{\mu \gamma}-p_{3 \mu}p_{3 \gamma}+M^2 g_{ \mu \gamma}\\
V_{\alpha \mu \gamma} (p_1, -k_1, p_3) p_3 ^\gamma =-[(p_1^2-M^2) g_{\alpha \mu}-p_{1\alpha} p_{1\mu}]-M^2 g_{\alpha \mu }\nonumber \\
p_1^{\alpha} V_{\alpha \nu \gamma} (p_1, -k_2, p_3)=(p_3^2-M^2)g_{\nu \gamma}-p_{3 \nu}p_{3 \gamma}+(M^2-M_Z^2) g_{ \nu \gamma}\\
V_{\alpha \nu \gamma} (p_1, -k_2, p_3) p_3 ^\gamma =-[(p_1^2-M^2) g_{\alpha \nu}-p_{1\alpha} p_{1\nu}]-(M^2-M_Z^2) g_{\alpha \nu }\nonumber
\end{eqnarray}
\begin{equation}
\label{a9}
p_1^{\alpha} V_{\alpha \mu \gamma} (p_1, -k_1, p_3) p_3 ^\gamma=0
\end{equation}
\begin{equation}
\label{a10}
p_1^{\alpha} V_{\alpha \nu \gamma} (p_1, -k_2, p_3) p_3 ^\gamma=-M_Z^2 p_{3\nu}=M_Z^2p_{1\nu}
\end{equation}

\end{document}